\documentclass[nojss]{jss}

\usepackage{amsmath}
\usepackage{amssymb}
\usepackage{booktabs}
\usepackage{listings}
\usepackage{subcaption}
\usepackage{tabularx}

\newcommand{\Connect}{\mathbb{C}}
\newcommand{\fatmu}{\boldsymbol{\mu}}
\newcommand{\fatnu}{\boldsymbol{\nu}}
\newcommand{\Intdom}{\mathbf{Z}}
\newcommand{\Ncompartments}{N_{\mbox{{\tiny comp}}}}
\newcommand{\Nconcentrations}{N_{\mbox{{\tiny conc}}}}
\newcommand{\Ntransitions}{N_{\mbox{{\tiny trans}}}}
\newcommand{\Nnodes}{N_{\mbox{{\tiny nodes}}}}
\newcommand{\Realdom}{\mathbf{R}}
\newcommand{\Stoich}{\mathbb{S}}
\newcommand{\X}{\mathbb{X}}
\newcommand{\Y}{\mathbb{Y}}

\author{Stefan Widgren\\National Veterinary Institute\\
  and Uppsala University\\Sweden\And
  Pavol Bauer\\Uppsala University\\Sweden\AND
  Robin Eriksson\\Uppsala University\\Sweden\And
  Stefan Engblom\\Uppsala University\\Sweden}

\Plainauthor{Stefan Widgren, Pavol Bauer, Robin Eriksson, Stefan
  Engblom}

\title{\pkg{SimInf}: An \proglang{R} package for Data-driven
  Stochastic Disease Spread Simulations}

\Plaintitle{SimInf: An R package for Data-Driven Stochastic Disease
  Spread Simulations}

\Shorttitle{\pkg{SimInf}: Data-driven Stochastic Disease Spread
  Simulations}

\Abstract{

  We present the \proglang{R} package \pkg{SimInf} which provides an
  efficient and very flexible framework to conduct data-driven
  epidemiological modeling in realistic large scale disease spread
  simulations.  The framework integrates infection dynamics in
  subpopulations as continuous-time Markov chains using the Gillespie
  stochastic simulation algorithm and incorporates available data such
  as births, deaths and movements as scheduled events at predefined
  time-points.  Using \proglang{C} code for the numerical solvers and
  \proglang{OpenMP} to divide work over multiple processors ensures
  high performance when simulating a sample outcome.  One of our
  design goal was to make \pkg{SimInf} extendable and enable usage of
  the numerical solvers from other \proglang{R} extension packages in
  order to facilitate complex epidemiological research.  In this
  paper, we provide a technical description of the framework and
  demonstrate its use on some basic examples.  We also discuss how to
  specify and extend the framework with user-defined models.

}

\Plainkeywords{computational epidemiology, discrete-event simulation,
  multicore implementation, stochastic modeling}

\Address{
  Stefan Widgren\\
  Department of Disease Control and Epidemiology\\
  National Veterinary Institute\\
  SE-751 89 Uppsala, Sweden\\
  E-mail: \email{stefan.widgren@sva.se}\\
  \textit{and}\\
  Division of Scientific Computing\\
  Department of Information Technology\\
  Uppsala University\\
  SE-751 05 Uppsala, Sweden\\
  E-mail: \email{stefan.widgren@it.uu.se}\\
  \\
  Pavol Bauer\\
  Division of Scientific Computing\\
  Department of Information Technology\\
  Uppsala University\\
  SE-751 05 Uppsala, Sweden\\
  E-mail: \email{pavol.bauer@it.uu.se}\\
  \\
  Robin Eriksson\\
  Division of Scientific Computing\\
  Department of Information Technology\\
  Uppsala University\\
  SE-751 05 Uppsala, Sweden\\
  E-mail: \email{robin.eriksson@it.uu.se}\\
  \\
  Stefan Engblom\\
  Division of Scientific Computing\\
  Department of Information Technology\\
  Uppsala University\\
  SE-751 05 Uppsala, Sweden\\
  E-mail: \email{stefane@it.uu.se}
}

\begin{document}



\section{Introduction}

Cattle can act as a reservoir for \textit{Salmonella} and
Verotoxin-producing \textit{Escherichia coli} (VTEC), two important
examples of zoonotic food-borne pathogens.  In order to develop
effective control strategies, it is necessary to understand the spread
of zoonotic diseases in the cattle population \citep{Newell2010}.
Since cattle are aggregated into spatially segregated farms, it is
natural to use a metapopulation framework and partition the cattle
population into interacting subpopulations \citep{Grenfell1997,
  Keeling2010}.  Furthermore, livestock data is commonly available,
with information on births, deaths and movements
\citep{BrooksPollock2014}.  Consequently, detailed spatiotemporal
demographic data and the transportation network are available and can
be used for epidemiologically relevant factors when simulating the
infection process within each subpopulation, coupled with spread among
subpopulations governed by spatial proximity and livestock movements.
However, incorporating large amounts of data in simulations is
computationally challenging and requires efficient algorithms.

In this work, we present the \proglang{R} \citep{R} package
\pkg{SimInf}, a flexible framework for data-driven spatio-temporal
disease spread modeling, designed to efficiently handle population
demographics and network data.  The framework integrates infection
dynamics in each subpopulation as continuous-time Markov chains (CTMC)
using the Gillespie stochastic simulation algorithm (SSA)
\citep{Gillespie1977} and incorporates available data such as births,
deaths or movements as scheduled events.  A scheduled event is used to
modify the state of a subpopulation at a predefined time-point.  Using
compiled \proglang{C} \citep{Kernighan1988} code, rather than
interpreted \proglang{R} code, for the numerical solvers ensures high
performance when simulating a model.  To further improve performance,
\proglang{OpenMP} \citep{OpenMP2008} is used to divide work over
multiple processors and perform computations in parallel.
Furthermore, the framework has a well-defined interface to incorporate
data that is shared among all subpopulations (global) and data that is
specific to each subpopulation (local), allowing sophisticated models
to be straightforwardly formulated.  The proposed approach was used to
study spread and control of VTEC in the complete Swedish cattle
population, incorporating almost ten years of scheduled events data in
a network of about 40,000 subpopulations \citep{Bauer2016,
  Widgren2016, Widgren2018}.  Even if development of \pkg{SimInf} was
inspired by livestock diseases and models driven by available data,
the design is of completely general character and applies to arbitrary
metapopulation models.  One of our design goal was to make
\pkg{SimInf} extendable and enable usage of the numerical solvers from
other \proglang{R} extension packages in order to facilitate complex
epidemiological research.  To support this, \pkg{SimInf} has
functionality to generate the required \proglang{C} and \proglang{R}
code from a model specification.

Various packages in the CRAN repository implement SSA to simulate a
continuous-time stochastic process.  The \pkg{GillespieSSA} package
\citep{Pineda-Krch2008}, on CRAN since 2007, but apparently not
actively maintained after 2012, implements both the direct method and
three approximate methods.  The \pkg{hybridModels} package
\citep{hybridModels} uses \pkg{GillespieSSA} internally to simulate
infections using a metapopulation model coupled with spread among
subpopulations. Since each outcome of a stochastic process is
different, it is (generally) necessary to study many realisations of
the process to see the distribution of outcomes consistent with the
model structure and parameterization.  Therefore, performance of the
simulator becomes critical when using these methods in an applied
context.  Because the algorithms in \pkg{GillespieSSA} are implemented
in \proglang{R}, the computational efficiency is limited in comparison
with implementations in a compiled language, for example, \proglang{C}
or \proglang{C++}. The \pkg{adaptivetau} package \citep{adaptivetau}
uses a hybrid \proglang{R}/\proglang{C++} strategy to implement the
direct method and adaptive tau leaping \citep{Cao2007}.

There exists several related \proglang{R} packages for epidemiological
analysis on the Comprehensive \proglang{R} Archive Network (CRAN).
For example, the package \pkg{amei} \citep{amei}, designed for finding
optimal intervention strategies to minimize total expected cost to
control a disease outbreak.  Another package is \pkg{surveillance}, a
framework for monitoring, modeling, and regression analysis of
infectious diseases, \citep{surveillance}.  \pkg{EpiModel}
\citep{EpiModel} is an \proglang{R} package that, includes a framework
for modeling spread of diseases on networks.  In \pkg{EpiModel} the
individual is the unit and transmission between individuals is
modelled through a contact network in discrete time.

The remainder of this paper is organized as follows.  In
\S\ref{sec:modeling} we summarize the mathematical foundation for our
framework. Section \S\ref{sec:framework} gives a technical description
of the simulation framework.  In \S\ref{sec:examples} we illustrate
the use of the package by some worked examples.  In \S\ref{sec:extend}
we demonstrate how to extend \pkg{SimInf} with user-defined models.
Finally, in \S\ref{sec:benchmark} we provide a small benchmark between
various \proglang{R} packages of the run-time to simulate SSA
trajectories.


\section{Epidemiological modeling}
\label{sec:modeling}

In mathematical modeling of the dynamics of an infectious disease in a
population, the population under study is commonly divided into
compartments representing discrete health states, together with
assumptions about the transition rates for individuals to move from
one compartment to another \citep{Kermack1927, Andersson1991,
  Keeling2007}. In order to capture spatial characteristics of a
disease process, a compartment model can be further partioned into
metapopulations (cities, households, farms), i.e., subpopulations with
its own infection dynamics \citep{Grenfell1997}.  Since the population
size of each subpopulation is small, it is often necessary to use
stochastic models, e.g., to account for the random event that an
infection will become extinct \citep{Bartlett1957}.  A stochastic
compartment model is naturally formulated as a CTMC using SSA to
simulate the number of individuals within each compartment through
time \citep{Keeling2007}.  Consequently, SSA is often used when
modeling various infectious diseases, for example, Ebola virus disease
outbreak \citep{King2015}, seasonality of influenza epidemics
\citep{Dushoff2004}, avian influenza virus in bird populations
\citep{Breban2009}, paratuberculosis infection in cattle
\citep{Smith2015}.

\begin{figure}
  \begin{center}
    \includegraphics{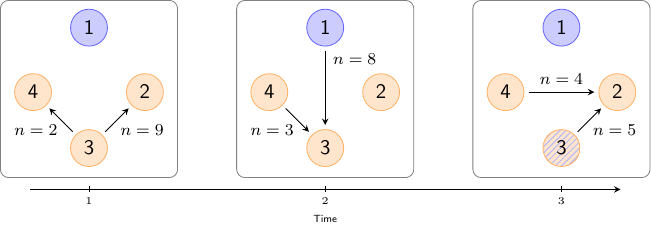}
  \end{center}
  \caption{Illustration of movements as a temporal network.
    Each time step depicts movements during one time unit, for
    example, a day.  The network has $N = 4$ nodes where node $1$ is
    infected and nodes $2$--$4$ are non-infected.  Arrows indicate
    movements of individuals from a source node to a destination node and
    labels denote the size of the shipment.  Here, infection may
    spread from node $1$ to node $3$ at $t=2$ and then from node $3$
    to node $2$ at $t=3$.
    \label{fig:temporal-network}}
\end{figure}

In order to model disease spread on a larger scale, the infection
process within each subpopulation must be coupled with spread among
subpopulations.  For example, livestock movements are an important
transmission route for many infectious diseases and can transfer
infectious individuals between farms over large distances
\citep{Danon2011}.  The livestock movements create complex dynamic
interactions among farms that can be represented as a directed
temporal network \citep{Kempe2002, Bajardi2011, Dutta2014}.  In
network terminology, each farm is represented by a \textit{node} (also
called a vertex).  Moreover, each movement forms an \textit{edge}
(also called a link) between two nodes and following all edges through
time, an infection may ``flow'' in the network and spread from node to
node. Let $\Nnodes$ denote the number of nodes in a population, and
let $i, j, k \in \{1, ..., \Nnodes\}$ denote three distinct nodes.  As
illustrated in Figure~\ref{fig:temporal-network}, just because there
exists an edge from $i$ to $j$ does not mean that there exists an edge
from $j$ to $i$.  Moreover, the existence of an edge from $i$ to $j$
and one from $j$ to $k$ does not imply there exists a path from $i$ to
$k$.  Furthermore, note that the order of the edges matter, consider
swapping the first and second time step in
Figure~\ref{fig:temporal-network}, then another path for spread is
possible, namely from node $3$ to node $4$ and then to node $2$.

In \S\ref{subsec:local}--\ref{subsec:numerics}, we provide an overview
of the epidemiological modeling framework employed in \pkg{SimInf}.
The overall approach consists of CTMCs as a general model of the
dynamics of the epidemiological state. Importantly, we also allow for
variables obeying ordinary differential equations (ODEs). For example,
this readily supports modeling infections that have an indirect
transmission route, e.g., via shedding of a pathogen to the
environment \citep{Ayscue2009, Breban2009}.  Additionally, the
framework handles externally defined demographic and movement events.
In \S\ref{subsec:local}--\ref{subsec:global} below we distinguish
between the \emph{local} dynamics that describes the evolution of the
epidemiological state at a single node, and the \emph{global}
dynamics, which describes the system at the network level. The overall
numerical approach underlying \pkg{SimInf} is described in
\S\ref{subsec:numerics}. We draw much of the material here from
\citep{Bauer2016, siminf_Ch}.

\subsection{Local dynamics}
\label{subsec:local}

We describe the state of a single node with a \emph{state vector}
$X(t) \in \Intdom_{+}^{\Ncompartments}$, which counts the number of
individuals at each of $\Ncompartments$ compartments at time $t$. The
transitions between these compartments are stochastic and are
described by the transition matrix $\Stoich \in
\Intdom^{\Ncompartments \times \Ntransitions}$ and the transition
intensities $R: \Intdom_{+}^{\Ncompartments} \to
\Realdom_{+}^{\Ntransitions}$, assuming $\Ntransitions$ different
transitions. We then form a \textit{random counting measure}
$\mu_{k}(dt) = \mu(R_{k}(X(t-)); \, dt)$ that is associated with a
Poisson process for the $k$th intensity $R_{k}(X(t-)$, which in turn
depends on the state prior to any transition at time $t$, that is,
$X(t-)$.

The local dynamics can then compactly be described by a pure jump
stochastic differential equation (SDE),
\begin{align}
  \label{eq:vectorJSDE}
  dX(t) &= \Stoich\fatmu(dt),
\end{align}
where $\fatmu(dt)$ is a vector measure built up from the scalar
counting measures $\fatmu(dt) = [\mu_{1}(dt),\ldots,$
  $\mu_{\Ntransitions}(dt)]^\top$. If at time $t$, transition $k$
occurs, then the state vector is updated according to
\begin{align}
  \label{eq:eventupdate}
   X(t)=X(t-)+\Stoich_k,
\end{align}
with $\Stoich_k$ the $k$th column of $\Stoich$. In
\eqref{eq:vectorJSDE} the $\Ntransitions$ different epidemiological
state transitions are competing in the sense of independent Poisson
processes. The `winning' process decides what event happens and
changes the state according to \eqref{eq:eventupdate}. The simulation
then proceeds under the Markov assumption where previous events are
remembered via the state variable $X$ only.

To make this abstract notation a bit more concrete we consider a
traditional example as follows. In an SIS-model the transitions
between a susceptible and an infected compartment can be written as
\begin{align}
\label{eq:sirtrans}
  &\left. \begin{array}{rl}
    S+I &\xrightarrow{\beta} 2I \\
    I &\xrightarrow{\gamma} S \\
  \end{array} \right\}.
  \intertext{With a state vector consisting of two compartments $X =
    [S,I]$, i.e., the number of susceptible and infected
    individuals, respectively, we can then write the transition matrix
    and intensity vector as}
  \label{eq:SIRstoich}
    \Stoich &= \left[ \begin{array}{rr}
        -1 & 1 \\
        1  & -1
      \end{array} \right], \\
    R(x) &= [\beta x_{1}x_{2},\gamma x_{2}]^\top.
\end{align}
To connect this with traditional ODE-based models, note that,
replacing the random measure in \eqref{eq:vectorJSDE} with its mean
drift, we arrive at
\begin{align}
  \label{eq:ODE}
  \frac{dx(t)}{dt} &= \Stoich R(x),
\end{align}
where now the state variable $x \in \Realdom^{\Ncompartments}$. The
differences between \eqref{eq:vectorJSDE} and \eqref{eq:ODE} are that
the randomness and discreteness of the state variable are not present
in the latter formulation. If these features are thought to be
important, then \eqref{eq:vectorJSDE} is an accurate stochastic
alternative to \eqref{eq:ODE}, relying only on the Markovian
``memoryless'' assumption.

There are, however, situations where we would like to mix the discrete
stochastic model with a concentration-type ODE model. In a multi-scale
description there are typically variables for which a continuous
description is more natural: a typical example is the concentration of
bacteria in an infectious environment for which individual counting
would clearly not be feasible.

Assuming an additional concentration state vector $Y \in
\Realdom^{\Nconcentrations}$ a general model which augments
\eqref{eq:vectorJSDE} is
\begin{align}
  \label{eq:JSDE_ODE}
  \left. \begin{array}{rcl}
    dX(t) &=& \Stoich\fatmu(dt) \\
    Y'(t) &=& f(X(t),Y(t)) \\
  \end{array} \right\},
\end{align}
where now the random measure depends also on the concentration
variable,
\begin{align}
  \fatmu(dt) = \fatmu(R(X(t-),Y(t)),dt).
\end{align}
The overall combined state vector is then $[X; \; Y] \in
[\Intdom^{\Ncompartments}; \; \Realdom^{\Nconcentrations}]$.

\subsection{Global dynamics}
\label{subsec:global}

To extend the local dynamics to a network model consisting of
$\Nnodes$ nodes we first define the overall state matrices $\X \in
\Intdom_{+}^{\Ncompartments \times \Nnodes}$ and $\Y \in
\Realdom^{\Nconcentrations \times \Nnodes}$ and then extend
\eqref{eq:JSDE_ODE} to
\begin{align}
  \label{eq:local1}
  d\X^{(i)}(t) &= \Stoich\fatmu^{(i)}(dt), \\
  \label{eq:local2}
  \frac{d\Y^{(i)}(t)}{dt} &= f(\X^{(i)},\Y^{(i)}),
\end{align}
where $i \in \{1,...,\Nnodes\}$ is the node index.

We then consider the $\Nnodes$ nodes being the vertices of an
undirected graph $\mathcal{G}$ with interactions defined in terms of
the counting measures $\fatnu^{(i,j)} = \fatnu^{(i,j)}(dt)$ and
$\fatnu^{(j,i)}$. Here $\fatnu^{(i,j)}$ represents the state changes
due to an inflow of individuals from node $i$ to node $j$, and
$\fatnu^{(j,i)}$ represents an inflow of individuals from node $j$ to
node $i$, assuming node $j$ being in the connected component $C(i)$ of
node $i$, and vice versa. We denote the connected components of the
graph $\mathcal{G}$ as the matrix
$\Connect \in \Intdom_{+}^{\Ncompartments \times \Nnodes}$.

The network dynamics is then written as
\begin{align}
  \label{eq:global1}
  d\X^{(i)}_{t} &= -\sum_{j \in C(i)} \Connect\fatnu^{(i,j)}(dt)+
  \sum_{j; \, i \in C(j)} \Connect\fatnu^{(j,i)}(dt), \\
  \label{eq:global2}
  \frac{d\Y^{(i)}(t)}{dt} &= -\sum_{j \in C(i)} g(\X^{(i)},\Y^{(i)})+
  \sum_{j; \, i \in C(j)} g(\X^{(j)},\Y^{(j)}).
\end{align}

In \eqref{eq:global2}, $g$ is similarly the ``flow'' of the
concentration variable $\Y$ between the nodes in the network. For
example, this could be the natural modeling target for concentration
variables $\Y$ which are transported via surface water or air.

Combining this with \eqref{eq:local1}--\eqref{eq:local2} we obtain the
overall dynamics
\begin{align}
  \label{eq:master1}
  d\X^{(i)}(t) &= \Stoich\fatmu^{(i)}(dt)-
  \sum_{j \in C(i)} \Connect\fatnu^{(i,j)}(dt)+
  \sum_{j; \, i \in C(j)} \Connect\fatnu^{(j,i)}(dt), \\
  \label{eq:master2}
  \frac{d\Y^{(i)}(t)}{dt} &= f(\X^{(i)},\Y^{(i)})-
  \sum_{j \in C(i)} g(\X^{(i)},\Y^{(i)})+
  \sum_{j; \, i \in C(j)} g(\X^{(j)},\Y^{(j)}).
\end{align}

Note that $\fatnu^{(i,j)}$ and $\fatnu^{(j,i)}$ may be equivalently
employed for externally scheduled events given by data using an
equivalent construction in terms of Dirac measures. This is the case,
for example, when intra-nodal transport data of individuals are
available.

\subsection{Numerical method}
\label{subsec:numerics}

In \pkg{SimInf}, we solve \eqref{eq:master1}--\eqref{eq:master2} by
splitting the local update scheme \eqref{eq:local1}--\eqref{eq:local2}
from the global update scheme
\eqref{eq:global1}--\eqref{eq:global2}. We discretize time as $0 =
t_{0} < t_{1} < t_2 < \cdots$, which is partially required as external
data has to be incorporated at some finitely resolved time stamps. The
numerical method of \pkg{SimInf} can then be written per node $i$ as
\begin{align}
  \label{eq:numstep1}
  \tilde{\X}_{n+1}^{(i)} &= \X_{n}^{(i)} + \int^{t_{n+1}}_{t_n}
  \Stoich \fatmu^{(i)}(ds), \\
  \label{eq:numstep2}
  \X_{n+1}^{(i)} &= \tilde{\X}^{(i)}_{n+1}-\int^{t_{n+1}}_{t_n} \sum_{j \in C(i)}
  \Connect\fatnu^{(i,j)}(ds)+\int^{t_{n+1}}_{t_n} \sum_{j; \, i \in C(j)}
  \Connect\fatnu^{(j,i)}(ds), \\
  \label{eq:numstep3}
  \Y_{n+1}^{(i)} &= \Y_{n}^{(i)} + f(\tilde{\X}_{n+1}^{(i)},
  \Y_{n}^{(i)}) \, \Delta t_{n} \\
  \nonumber
  &\phantom{=}
  -\sum_{j \in C(i)} g(\tilde{\X}_{n+1}^{(i)},\Y_{n}^{(i)})\Delta t_{n}+
  \sum_{j; \, i \in C(j)} g(\tilde{\X}_{n+1}^{(j)},\Y_{n}^{(j)})\Delta t_{n}.
\end{align}

In this scheme, \eqref{eq:numstep1} forms the local stochastic step,
that is in practice simulated by the Gillespie method
\citep{Gillespie1977}. Equation~\eqref{eq:numstep2} is the data step,
where externally scheduled events are incorporated. Note that the
stochastic step evolves in continuous time in the interval
$[t_n,t_{n+1}]$, and the data step operates only on the final state
$\tilde{\X}$ at $t_{n+1}$. The final step \eqref{eq:numstep3} is just
the Euler forward method in time with time-step
$\Delta t_{n} = t_{n+1}-t_{n}$ for the concentration variable $\Y$.


\begin{figure}
  \begin{center}
    \includegraphics{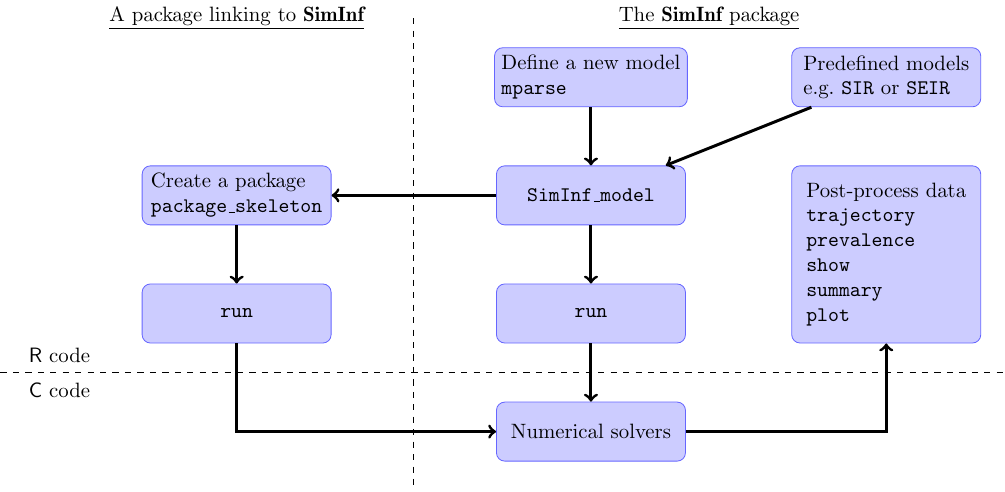}
  \end{center}
  \caption{Schematic overview of the functionality in
    \pkg{SimInf}. The central object is the S4 class
    \code{Simlnf\_model} which contains the specification and data for
    a model.  A new model object is created using \code{mparse} or a
    predefined template, for example, \code{SIR} or \code{SEIR}.  A
    stochastic trajectory is simulated from a model using \code{run}.
    For computational efficiency, the numerical solvers are
    implemented in \proglang{C} code.  There are several functions in
    \pkg{SimInf} to facilitate analysis and post-processing of
    simulated data, for example, \code{trajectory}, \code{prevalence}
    and \code{plot}.  \pkg{SimInf} supports usage of the numerical
    solvers from other \proglang{R} packages via the \emph{LinkingTo}
    feature in \proglang{R}. \label{fig:overview}}
\end{figure}

\section{Technical description of the simulation framework}
\label{sec:framework}

The overall design of \pkg{SimInf} was inspired and partly adapted
from the Unstructured Mesh Reaction-Diffusion Master Equation (URDME)
framework \citep{Engblom2009, Drawert2012}.  \pkg{SimInf} uses object
oriented programming and the \code{S4} class \citep{Chambers2008}
\code{SimInf\_model} is central and provides the basis for the
framework.  A \code{SimInf\_model} object supplies the state-change
matrix, the dependency graph, the scheduled events, and the initial
state of the system.  Briefly, the state-change matrix defines the
effect of the disease transitions on the state of the system while the
dependency graph indicates the transition rates that need to be
updated after a given disease transition.  Additionally, model
specific code written in \proglang{C} specifies the transition rate
functions for the disease transitions in the system.  All predefined
models in \pkg{SimInf} have a generating function
\citep{Chambers2008}, with the same name as the model, to initialize
the data structures for that specific model, see the examples in
\S\ref{sec:examples}.  A model can also be created from a model
specification using the \code{mparse} method, further described in
\S\ref{sec:extend}.  After a model is created, a simulation is started
with a call to the \code{run} method and if execution is successful,
it returns a modified \code{SimInf\_model} object with a single
stochastic solution trajectory attached to it.  \pkg{SimInf} provides
several utility functions to inspect simulated data, for example,
\code{show}, \code{summary} and \code{plot}.  To facilitate custom
analysis, \pkg{SimInf} provides the \code{trajectory} and
\code{prevalence} methods that both return a \code{data.frame} with
simulated data.  Figure~\ref{fig:overview} shows a schematic overview
of the functionality in \pkg{SimInf}.  The overall modular design
makes extensions easy to handle.

\subsection{Installation}
\label{sec:installation}

The most recent stable version of \pkg{SimInf} is distributed via CRAN
(\url{https://CRAN.R-project.org/package=SimInf}) and may, depending
on your platform, be available in source form or compiled binary form.
The development version is available on GitHub
(\url{https://github.com/stewid/SimInf}).  A binary form of
\pkg{SimInf} for macOS or Windows can be installed directly from CRAN.
However, if you install \pkg{SimInf} from source (from CRAN or a
\code{.tar.gz} file), the installation process requires a \proglang{C}
compiler, and that the GNU Scientific Library (GSL)
\citep{Gallassi2009} is installed on your system and is on the path.
On a Windows machine you first need to download and install Rtools
from \url{https://cran.r-project.org/bin/windows/Rtools}.  Note that
GSL (\url{https://www.gnu.org/software/gsl/}) is not an \proglang{R}
add-on package, but needs to be installed separately, for example,
from a terminal using: \code{'sudo apt-get install libgsl0-dev'} on
Debian and Ubuntu, \code{'sudo yum install gsl-devel'} on Fedora,
CentOS or RHEL, or \code{'brew install gsl'} on macOS with the
Homebrew package manager.  On Windows, the GSL files are downloaded,
if needed, from \url{https://github.com/rwinlib/gsl} during the
installation of \pkg{SimInf}.  Furthermore, when you install
\pkg{SimInf} from source, depending on features of the compiler, the
package is compiled with support for \proglang{OpenMP}.  To find out
more about installing \proglang{R} add-on packages in general, the
\texttt{'R Installation and Administration'}
(\url{https://cran.r-project.org/manuals.html}) manual describes the
process in detail.  After installing the package

\begin{Schunk}
\begin{Sinput}
R> install.packages("SimInf")
\end{Sinput}
\end{Schunk}

it is loaded in \proglang{R} with the following command

\begin{Schunk}
\begin{Sinput}
R> library("SimInf")
\end{Sinput}
\end{Schunk}

\begin{table}
  \small
  \begin{tabularx}{\textwidth}{l X}
    \toprule
    Slot & Description\\
    \midrule

    \code{S} & Each column corresponds to a state transition, and
    execution of state transition $j$ amounts to adding the \code{S[,
        j]} column to the state vector \code{u[, i]} of node $i$ where
    the transition occurred.  Sparse matrix ($\Ncompartments \times
    \Ntransitions$) of object class \code{dgCMatrix}.\\

    \code{G} & Dependency graph that indicates the transition rates
    that need to be updated after a given state transition has
    occurred.  A non-zero entry in element \mbox{\code{G[i, j]}}
    indicates that transition rate $i$ needs to be recalculated if the
    state transition $j$ occurs.  Sparse matrix ($\Ntransitions \times
    \Ntransitions$) of object class \code{dgCMatrix}.\\

    \code{tspan} & A vector of increasing time points where the state
    of each node is to be returned.\\

    \code{U} & The result matrix with the number of individuals in
    each compartment in every node.  \mbox{\code{U[, j]}} contains the
    number of individuals in each compartment at \code{tspan[j]}.
    \mbox{\code{U[1:$\Ncompartments$, j]}} contains the number of
    individuals in each compartment in node $1$ at \code{tspan[j]}.
    \mbox{\code{U[($\Ncompartments$ + 1):(2 * $\Ncompartments$), j]}}
    contains the number of individuals in each compartment in node $2$
    at \code{tspan[j]} etc.  Integer matrix ($\Nnodes \Ncompartments
    \times \text{length}(\text{tspan})$).\\

    \code{U\_sparse} & It is possible to run the simulator and write
    the number of individuals in each compartment to the
    \code{U\_sparse} sparse matrix (\code{dgCMatrix}), which can save
    a lot of memory if the model contains many nodes and time-points,
    but where only a few of the data points are of interest.  If
    \code{U\_sparse} is non-empty when \code{run} is called, the
    non-zero entries in \code{U\_sparse} indicates where the number of
    individuals should be written to \code{U\_sparse}.  The layout of
    the data in \code{U\_sparse} is identical to \code{U}.  Please
    note that the data in \code{U\_sparse} is numeric and that the
    data in \code{U} is integer.\\

    \code{u0} & The initial number of individuals in each compartment
    in every node.  Integer matrix ($\Ncompartments \times
    \Nnodes$).\\

    \code{V} & The result matrix for the real-valued continuous
    state.  \mbox{\code{V[, j]}} contains the real-valued state of the
    system at \code{tspan[j]}.  Numeric matrix ($\Nnodes N_{ld} \times
    \text{length}(\text{tspan})$).\\

    \code{V\_sparse} & It is possible to run the simulator and write
    the real-valued continuous state to the \code{V\_sparse} sparse
    matrix (\code{dgCMatrix}), which can save a lot of memory if the
    model contains many nodes and time-points, but where only a few of
    the data points are of interest.  If \code{V\_sparse} is non-empty
    when \code{run} is called, the non-zero entries in
    \code{V\_sparse} indicates where the real-valued continuous state
    should be written to \code{V\_sparse}.  The layout of the data in
    \code{V\_sparse} is identical to \code{V}.\\

    \code{v0} & The initial value for the real-valued continuous
    state.  Numeric matrix ($N_{ld} \times \Nnodes$).\\

    \code{ldata} & A numeric matrix with local data specific to each
    node.  The column \code{ldata[, j]} contains the local data vector
    for node $j$.  The local data vector is passed as an argument to
    the transition rate functions and the post time step function.\\

    \code{gdata} & A numeric vector with global data that is common to
    all nodes.  The global data vector is passed as an argument to the
    transition rate functions and the post time step function.\\

    \code{events} & Scheduled events to modify the discrete state of
    individuals in a node at a pre-defined time $t$.  \code{S4} class
    \code{SimInf\_events}, see \S\ref{sec:events} and
    Table~\ref{table:scheduled:events}.\\

    \code{C\_code} & Character vector with optional model \proglang{C}
    code, see \S\ref{sec:extend}.  If non-empty, the \proglang{C} code
    is written to a temporary file when the \code{run} method is
    called.  The temporary file is compiled and the resulting DLL is
    dynamically loaded.  The DLL is unloaded and the temporary files
    are removed after running the model.\\

    \bottomrule
  \end{tabularx}
  \caption{Description of the slots in the S4 class
    \code{SimInf\_model} that defines the epidemiological model.
    $\Ntransitions$ is the number of state transitions in the model.
    $\Ncompartments$ is the number of compartments in the model.
    $\Nnodes$ is the number of nodes in the model.  $N_{ld}$ is the
    number of local data specific to each node and equals
    \code{dim(ldata)[1]}.}
  \label{table:SimInf:model}
\end{table}

\subsection{Specification of an epidemiological model}
\label{sec:model}

The within-node disease spread model in \pkg{SimInf} is specified as a
compartment model with the individuals divided into compartments
defined by discrete disease statuses.  The model is defined by the
slots in the \code{S4} class \code{SimInf\_model}
(Table~\ref{table:SimInf:model}).  The compartments contains the
number of individuals in each of the $\Ncompartments$ disease states
in every $\Nnodes$ nodes.

Equation~\eqref{eq:numstep3}, the stochastic step, contains
$\Ntransitions$ state transitions and is processed using the two slots
\code{S} and \code{G}.  The \code{S} slot is the state-change matrix
($\Ncompartments \times \Ntransitions$) that determines how to change
the number of individuals in the compartments of a node when the
$j^{th}$ state transition occurs, where $1 \le j \le \Ntransitions$.
Each row corresponds to one compartment and each column to a state
transition.  Let \code{u[, i]} be the number of individuals in each
compartment in node $i$ at time $t_i$.  To move simulation time
forward in node $i$ to $t_i = t_i + \tau_i$, the vector \code{u[, i]}
is updated according to the $j^{th}$ transition by adding the
state-change vector \code{S[, j]} to \code{u[, i]}.  After updating
\code{u[, i]}, the transition rates must be recalculated to obtain the
time to the next event.  However, a state transition might not need
all transition rates to be recalculated.  The dependency graph
\code{G} is a matrix ($\Ntransitions \times \Ntransitions$) that
determines which transition rates that need to be recalculated.  A
non-zero entry in element \code{G[k, j]} indicates that transition
rate \code{k} needs to be recalculated if the $j^{th}$ state
transition occurs, where \mbox{$1 \le$ \code{k} $\le \Ntransitions$}.
Furthermore, the final step \eqref{eq:numstep3} is incorporated using
a model specific post time step callback to allow update of the
concentration variable $\Y$.

\begin{table}
  \small
  \begin{tabularx}{\textwidth}{l X}
    \toprule Slot & Description\\ \midrule

    \code{E} & Each row corresponds to one compartment in the model.
    The non-zero entries in a column indicate which compartments to
    sample individuals from when processing an event.  Which column to
    use for each event is specified by the \code{select} vector (see
    below).  \code{E} is a sparse matrix of class \code{dgCMatrix}.\\

    \code{N} & Determines how individuals in \textit{internal
      transfer} and \textit{external transfer} events are shifted to
    enter another compartment.  Each row corresponds to one
    compartment in the model.  The values in a column are added to the
    current compartment of sampled individuals to specify the
    destination compartment, for example, a value of \code{1} in an
    entry means that sampled individuals in this compartment are moved
    to the next compartment.  Which column to use for each event is
    specified by the \code{shift} vector (see below).  \code{N} is an
    integer matrix.\\

    \code{event} & Four event types are supported by the current
    solvers: \textit{exit}, \textit{enter}, \textit{internal
      transfer}, and \textit{external transfer}.  When assigning the
    events from a \code{data.frame}, they can either be coded as a
    numerical value or a character string: \textit{exit;} \code{0} or
    \code{'exit'}, \textit{enter;} \code{1} or \code{'enter'},
    \textit{internal transfer;} \code{2} or \code{'intTrans'}, and
    \textit{external transfer;} \code{3} or \code{'extTrans'}.
    Internally in \pkg{SimInf}, the event type is coded as a numerical
    value.\\

    \code{time} & Time of when the event occurs i.e., the event is
    processed when time is reached in the simulation.  \code{time} is
    an integer vector.\\

    \code{node} & The node that the event operates on.  Also the
    source node for an \textit{external transfer} event.  \code{node}
    is an integer vector, where $1 \le$ \code{node[i]} $\le
    \Nnodes$.\\

    \code{dest} & The destination node for an \textit{external
      transfer} event i.e., individuals are moved from \code{node} to
    \code{dest}, where $1 \le$ \code{dest[i]} $\le \Nnodes$.  Set
    \code{event = 0} for the other event types.  \code{dest} is an
    integer vector.\\

    \code{n} & The number of individuals affected by the event.
    \code{n} is an integer vector, where \code{n[i]} $\ge 0$.\\

    \code{proportion} & If \code{n[i]} equals zero, the number of
    individuals affected by \code{event[i]} is calculated by summing
    the number of individuals in the compartments determined by
    \code{select[i]} and multiplying with \code{proportion[i]}.
    \code{proportion} is a numeric vector, where $0 \le$
    \code{proportion[i]} $\le 1$.\\

    \code{select} & To process an \code{event[i]}, the compartments
    affected by the event are specified with \code{select[i]} together
    with the matrix \code{E}, where \code{select[i]} determines which
    column in \code{E} to use.  The specific individuals affected by
    the event are proportionally sampled from the compartments
    corresponding to the non-zero entries in the specified column in
    \code{E[, select[i]]}, where \code{select} is an integer vector.\\

    \code{shift} & Determines how individuals in \textit{internal
      transfer} and \textit{external transfer} events are shifted to
    enter another compartment.  The sampled individuals are shifted
    according to column \code{shift[i]} in matrix \code{N} i.e.,
    \code{N[, shift[i]]}, where \code{shift} is an integer vector.
    See above for a description of \code{N}.\\

    \bottomrule
  \end{tabularx}
  \caption{Description of the slots in the S4 class
    \code{SimInf\_events} that holds data to process scheduled events
    to modify the discrete state of individuals in a node at a
    pre-defined time $t$.  Each index, \code{i}, of the vectors
    represent one event.  $\Nnodes$ is the number of nodes in the
    model.}
  \label{table:scheduled:events}
\end{table}

Model-specific data that is passed to the transition-rate functions
and the post time-step function are stored in the two slots
\code{ldata} and \code{gdata} in the \code{SimInf\_model} object.  The
\code{ldata} matrix holds local data for each node where \code{ldata[,
  i]} is the data vector for node $i$.  Data that is global, i.e.,
shared between nodes, is stored in the \code{gdata} vector.

The \code{events} slot in the \code{SimInf\_model} holds data to
process the scheduled events, further described in \S\ref{sec:events}.

During simulation of one trajectory, the state of the system is
written to the two matrices \code{U} and \code{V}.  This happens at
each occasion the simulation time passes a time point in \code{tspan},
a vector of increasing time points.  The first and last element in
\code{tspan} determines the start- and end-point of the simulation.
The column \code{U[, m]} contains the number of individuals in each
compartment in every node at \code{tspan[m]}, where $1 \le$ \code{m}
$\le$ \code{length(tspan)}.  The first $\Ncompartments$ rows in
\code{U} contains the compartments of the first node.  The next
$\Ncompartments$ rows contains the compartments of the second node
etc.  The \code{V} matrix contains output from continuous state
variables.  The column \code{V[, m]} contains the values at
\code{tspan[m]}.  The rows are grouped per node and the number of rows
per node is determined by the number of continuous state variables in
that specific model.  It is also possible to configure the simulator
to write the state of the system to the sparse matrices
\code{U\_sparse} and \code{V\_sparse}, which can save a lot of memory
if the model contains many nodes and time-points, but where only a few
of the data points are of interest.  In order to use this feature,
call the \code{U} and \code{V} methods (before running a trajectory)
with a \code{data.frame} that specify the nodes, time-points and
compartments where the simulator should write the state of the system.
The initial state in each node is specified by the two matrices
\code{u0} and \code{v0} where \code{u0[, i]} is the initial number of
individuals in each compartment at node $i$ and \code{v0[, i]} is the
initial continuous state in node $i$.

\begin{table}
  \small
  \begin{tabularx}{\textwidth}{l X}
    \toprule
    Argument & Description\\
    \midrule
    \code{v\_new} & If a continuous state vector is used by a model,
    this is the new continuous state vector in the node after the post
    time step.  Exists only in \code{PTSFun}.\\
    \code{u} & The compartment state vector in the node.\\
    \code{v} & The current continuous state vector in the node.\\
    \code{ldata} & The local data vector for the node.\\
    \code{gdata} & The global data vector that is common to all nodes.\\
    \code{node} & The node index.  Note the node index is zero-based,
    i.e., the first node is $0$.\\
    \code{t} & Current time in the simulation.\\
    \bottomrule
  \end{tabularx}
  \caption{Description of the arguments to the transition rate
    functions (\code{TRFun}) and the post time step function
    (\code{PTSFun}).}
  \label{table:arguments}
\end{table}

\subsection{Specification of scheduled events}
\label{sec:events}

The scheduled events are used to modify the discrete state of
individuals in a node at a pre-defined time $t$.  There are four
different types of events; \textit{enter}, \textit{internal transfer},
\textit{external transfer} and \textit{exit}. The \textit{enter} event
adds individuals to a node, for example, due to births. The
\textit{internal transfer} event moves individuals between
compartments within one node.  For example, to simulate vaccination
and move individuals to a vaccinated compartment (see example in
\S\ref{sec:mparse-scheduled-events}).  Or ageing according to birth
records in an age-structured model \citep{Widgren2016}.  The
\textit{external transfer} event moves individuals from compartments
in one node to compartments in a destination node. Finally, the
\textit{exit} event removes individuals from a node, for example, due
to death. The event types are classified into those that operate on
the compartments of a single node $E_1 = \{\text{\textit{enter},
  \textit{internal transfer}, \textit{exit}}\}$ and those that operate
on the compartments of two nodes $E_2 = \{\text{\textit{external
    transfer}}\}$.  The parallel algorithm processes these two classes
of events differently, see Appendix~\ref{sec:pseudo-code} for
pseudo-code of the core simulation solver.  The scheduled events are
processed when simulation time reaches the time for any of the events.
Events that are scheduled at the same time are processed in the
following order: \textit{exit}, \textit{enter}, \textit{internal
  transfer} and \textit{external transfer}.

The S4 class \code{SimInf\_events} contains slots with data structures
to process events (Table~\ref{table:scheduled:events}).  The slots
\code{event}, \code{time}, \code{node}, \code{dest}, \code{n},
\code{proportion}, \code{select} and \code{shift}, are vectors of
equal length.  These vectors hold data to process one event: \code{e},
where $1 <$ \code{e} $\leq$ \code{length(event)}.  The event type and
the time of the event are determined by \code{event[e]} and
\code{time[e]}, respectively.  The compartments that \code{event[e]}
operates on, are specified by \code{select[e]} together with the slot
\code{E}.  Each row $\{1, 2, ..., \Ncompartments\}$ in the sparse
matrix \code{E}, represents one compartment in the model.  Let \code{s
  <- select[e]}, then each non-zero entry in the column \code{E[, s]}
includes that compartment in the \code{event[e]} operation.  The
definitions of all of these operations are a bit involved and to
quickly get an overview, schematic diagrams illustrating all of them
have been prepared, we refer to
Figures~\ref{fig:exit},\ref{fig:enter},\ref{fig:external},\ref{fig:internal},\ref{fig:external:shift}
in Appendix~\ref{sec:illustration-scheduled-events}.

\subsubsection{Processing of an enter event}

The \textit{enter} event adds \code{n[e]} individuals to one
compartment at \code{node[e]}, where the compartment is specified by a
non-zero entry in the row for the compartment in column \code{E[, s]}.
Please note that, if the column \code{E[, s]} contains several
non-zero entries, the individuals are added to the compartment
represented by the first non-zero row in column \code{E[, s]}.  The
values of \code{dest[e]}, \code{proportion[e]} and \code{shift[e]},
described below, are not used when processing an \textit{enter} event.
See Figure~\ref{fig:enter} in
Appendix~\ref{sec:illustration-scheduled-events} for an illustration
of a scheduled \textit{enter} event.

\subsubsection{Processing of an internal transfer event}

The \textit{internal transfer} event moves \code{n[e]} individuals
into new compartments within \code{node[e]}.  However, if \code{n[e]}
equals zero, the number of individuals to move is calculated by
multiplying the \code{proportion[e]} with the total number of
individuals in the compartments represented by the non-zero entries in
column \mbox{\code{E[, s]}}.  The individuals are then proportionally
sampled and removed from the compartments specified by \code{E[, s]}.
The next step is to move the sampled individuals to their new
compartment using the matrix \code{N} and \code{shift[e]}, where
\code{shift[e]} specifies which column in \code{N} to use.  Each row
$\{1, 2, ..., \Ncompartments\}$ in \code{N}, represents one
compartment in the model and the values determine how to move sampled
individuals before adding them to \code{node[e]} again.  Let \code{q
  <- shift[e]}, then each non-zero entry in \code{N[, q]} defines the
number of rows to move sampled individuals from that compartment i.e.,
sampled individuals from compartment \code{p} are moved to compartment
\code{N[p, q] + p}, where $1 \leq$ \code{N[p, q] + p} $\le
\Ncompartments$.  The value of \code{dest[e]}, described below, is not
used when processing an \textit{internal transfer} event.  See
Figure~\ref{fig:internal} in
Appendix~\ref{sec:illustration-scheduled-events} for an illustration
of a scheduled \textit{internal transfer} event.

\subsubsection{Processing of an external transfer event}

The \textit{external transfer} event moves individuals from
\code{node[e]} to \code{dest[e]}.  The sampling of individuals from
\code{node[e]} is performed in the same way as for an \textit{internal
  transfer} event.  The compartments at \code{node[e]} are updated by
subtracting the sampled individuals while adding them to the
compartments at \code{dest[e]}.  The sampled individuals are added to
the same compartments in \code{dest[e]} as in \code{node[e]}, unless
\code{shift[e]} $> 0$.  In that case, the sampled individuals change
compartments according to \code{N} as described in processing an
\textit{internal transfer} event before adding them to \code{dest[e]}.
See Figures \ref{fig:external} and \ref{fig:external:shift} in
Appendix~\ref{sec:illustration-scheduled-events} for illustrations of
scheduled \textit{external transfer} events.

\subsubsection{Processing of an exit event}

The \textit{exit} event removes individuals from \code{node[e]}.  The
sampling of individuals from \code{node[e]} is performed in the same
way as for an \textit{internal transfer} event.  The compartments at
\code{node[e]} are updated by subtracting the sampled individuals.
The values of \code{dest[e]} and \code{shift[e]} are not used when
processing an \textit{exit} event.  See Figure~\ref{fig:exit} in
Appendix~\ref{sec:illustration-scheduled-events} for an illustration
of a scheduled \textit{exit} event.

\subsection{Core simulation solvers}

The \pkg{SimInf} package uses the ability to interface compiled code
from R \cite{Chambers2008}.  The solvers are implemented in the
compiled language \proglang{C} and is called from \proglang{R} using
the \code{.Call()} interface \citep{Chambers2008}.  Using \proglang{C}
code rather than interpreted \proglang{R} code ensures high
performance when running the model.  To improve performance further,
the numerical solvers use \proglang{OpenMP} to divide work over
multiple processors and perform computations in parallel.  Two
numerical solvers are currently supported.  The default solver is a
split-step method named \code{'ssm'}, that uses direct SSA, but once
every unit of time, it also processes scheduled events and calls the
post time step function.  The other solver implements an \textit{``all
  events method''} \citep{Bauer2015} and is named \code{'aem'}.
Similarly, it also processes scheduled events and calls the post time
step function once every unit of time.  A core feature of the
\code{'aem'} solver is that transition events are carried out in
channels which access private streams of random numbers, in contrast
to the \code{'ssm'} solver where one uses only one stream for all
events.

\subsubsection{Function pointers}

The flexibility of the solver is partly achieved by using function
pointers \citep{Kernighan1988}.  A function pointer is a variable that
stores the address of a function that can be used to invoke the
function.  This provides a simple way to incorporate model specific
functionality into the solver.  A model must define one transition
rate function for each state transition in the model.  These functions
are called by the solver to calculate the transition rate for each
state transition in each node.  The output from the transition rate
function depends only on the state of the system at the current time.
However, the output is unique to a model and data are for that reason
passed on to the function for the calculation.  Furthermore, a model
must define the post time step function.  This function is called once
for each node each time the simulation of the continuous-time Markov
chain reaches the next day (or, more generally, the next unit of time)
and after the $E_1$ and $E_2$ events have been processed.  The main
purpose of the post time step function is to allow for a model to
update continuous state variables in each node.

The transition rate function is defined by the data type \code{TRFun}
and the post time step function by the data type \code{PTSFun}.  These
data types are defined in the header file \texttt{'src/SimInf.h'} and
shown below.  The arguments \code{v\_new}, \code{u}, \code{v},
\code{ldata}, \code{gdata}, \code{node}, and \code{t} of the functions
are described in Table~\ref{table:arguments}.

\begin{minipage}{\linewidth}
\begin{scriptsize}
\begin{lstlisting}[language=C]
  typedef double (*TRFun)(const int *u, const double *v, const double *ldata,
                          const double *gdata, double t);

  typedef int (*PTSFun)(double *v_new, const int *u, const double *v,
                        const double *ldata, const double *gdata,
                        int node, double t);
\end{lstlisting}
\end{scriptsize}
\end{minipage}

\subsubsection{Overview of the solvers}

Here follows an overview of the steps a solver performs to run a
trajectory, see Appendix~\ref{sec:pseudo-code} for pseudo-code for the
\code{'ssm'} solver and \texttt{'src/solvers/ssm/SimInf\_solver.c'}
for the source code.  The simulation starts with a call to the
\code{run} method with the model as the first argument and optionally
the number of threads to use.  This method will first call the
validity method on the model to perform error-checking and then call a
model specific \proglang{C} function to initialize the function
pointers to the transition rate functions and the post time step
function of the model.  Subsequently, the simulation solver is called
to run one trajectory using the model specific data, the transition
rate functions, and the post time step function.  If the
\code{C\_code} slot is non-empty, the \proglang{C} code is written to
a temporary file when the \code{run} method is called.  The temporary
file is compiled using \code{'R CMD SHLIB'} and the resulting DLL is
dynamically loaded.  The DLL is unloaded and the temporary files are
removed after running the model.  This is further described in
\S\ref{sec:extend}.

The solver simulates the trajectory in parallel if \proglang{OpenMP}
is available.  The default is to use all available threads.  However,
the user can specify the number of threads to use.  The solver divides
data for the $\Nnodes$ nodes and the $E_1$ events over the number
threads.  All $E_1$ events that affect node $i$ is processed in the
same thread as node $i$ is simulated in.  The $E_2$ events are
processed in the main thread.

The solver runs the continuous-time Markov chain for each node $i$.
For every time step $\tau_i$, the count in the compartments at node
$i$ is updated according to the state transition that occurred
(\S\ref{sec:model}).  The time to the next event is computed, after
recalculating affected transition rate functions (\S\ref{sec:model}).
When simulated time reaches the next day in node $i$ the $E_1$ events
are processed for that node (\S\ref{sec:events}).  The $E_2$ events
are processed when all nodes reaches the next day
(\S\ref{sec:events}).  Thereafter, the post time step function is
called to allow the model to incorporate model specific actions.  When
simulated time passes the next time in \code{tspan}, the count of the
compartments and the continuous state variables are written to
\code{U} and \code{V}.


\section[Model construction and data analysis: Basic examples]{%
  Model construction and data analysis: Basic examples}
\label{sec:examples}

\subsection[A first example: The SIR model]{%
  A first example: The \code{SIR} model}

\subsubsection[Specification of the SIR model without scheduled events]{%
  Specification of the \code{SIR} model without scheduled events}

This section illustrates the specification of the predefined
\code{SIR} model, which contains the three compartments susceptible
(\code{S}), infected (\code{I}) and recovered (\code{R}).  The
transmission route of infection to susceptible individuals is through
direct contact between susceptible and infected individuals.  The
\code{SIR} model has two state transitions in each node $i$,
\begin{align}
\label{eq:SIR}
\begin{array}{rcl}
  S_i & \xrightarrow{\beta S_i I_i / (S_i+I_i+R_i)} & I_i, \\
  I_i &\xrightarrow{\gamma I_i} & R_i,
  \end{array}
\end{align}
where $\beta$ is the transmission rate and $\gamma$ is the recovery
rate.  To create an \code{SIR} model object, we need to define
\code{u0}, a \code{data.frame} with the initial number of individuals
in each compartment when the simulation starts.  Let us consider a
node with 999 susceptible, 1 infected and 0 recoverd individuals.
Since there are no between-node interactions in this example, the
stochastic process in one node does not affect any other nodes in the
model.  Consequently, it is straightforward to run many realizations
of this model, simply by replicating a node in \code{u0}, for example,
\code{n = 1000} times.

\begin{Schunk}
\begin{Sinput}
R> n <- 1000
R> u0 <- data.frame(S = rep(999, n), I = rep(1, n), R = rep(0, n))
\end{Sinput}
\end{Schunk}

Next, we define the time period over which we want to simulate the
disease spread.  This is a vector of integers in units of time or a
vector of dates.  You specify those time points that you wish the
model to return results for.  The model itself does not run in
discrete time steps, but in continuous time, so this does not affect
the internal calculations of disease transitions through time.  In
this example we will assume that the unit of time is one day and
simulate over 180 days returning results every $7^{th}$ day.

\begin{Schunk}
\begin{Sinput}
R> tspan <- seq(from = 1, to = 180, by = 7)
\end{Sinput}
\end{Schunk}

We are now ready to create an \code{SIR} model and then use the
\code{run()} routine to simulate data from it.  For reproducibility,
we first call the \code{set.seed()} function and also specify the
number of threads to use for the simulation.  To use all available
threads, you only have to remove the threads argument.

\begin{Schunk}
\begin{Sinput}
R> model <- SIR(u0 = u0, tspan = tspan, beta = 0.16, gamma = 0.077)
R> set.seed(123)
R> result <- run(model = model, threads = 1)
\end{Sinput}
\end{Schunk}

The return value from \code{run()} is a \code{SimInf_model} object
with a single stochastic solution trajectory attached to it.  The
\code{show()} method of the \code{SimInf_model} class prints some
basic information about the model, such as the global data parameters
and the extremes, the mean and the quartiles of the count in each
compartment across all nodes.

\begin{Schunk}
\begin{Sinput}
R> result
\end{Sinput}
\begin{Soutput}
Model: SIR
Number of nodes: 1000
Number of transitions: 2
Number of scheduled events: 0

Global data
-----------
 Parameter Value
 beta      0.160
 gamma     0.077

Compartments
------------
    Min. 1st Qu. Median  Mean 3rd Qu.  Max.
 S 108.0   368.0  993.0 755.4   999.0 999.0
 I   0.0     0.0    1.0  30.4    38.0 235.0
 R   0.0     1.0    5.0 214.2   484.0 891.0
\end{Soutput}
\end{Schunk}

\begin{figure}
  \begin{center}
  \includegraphics[width=0.49\linewidth]{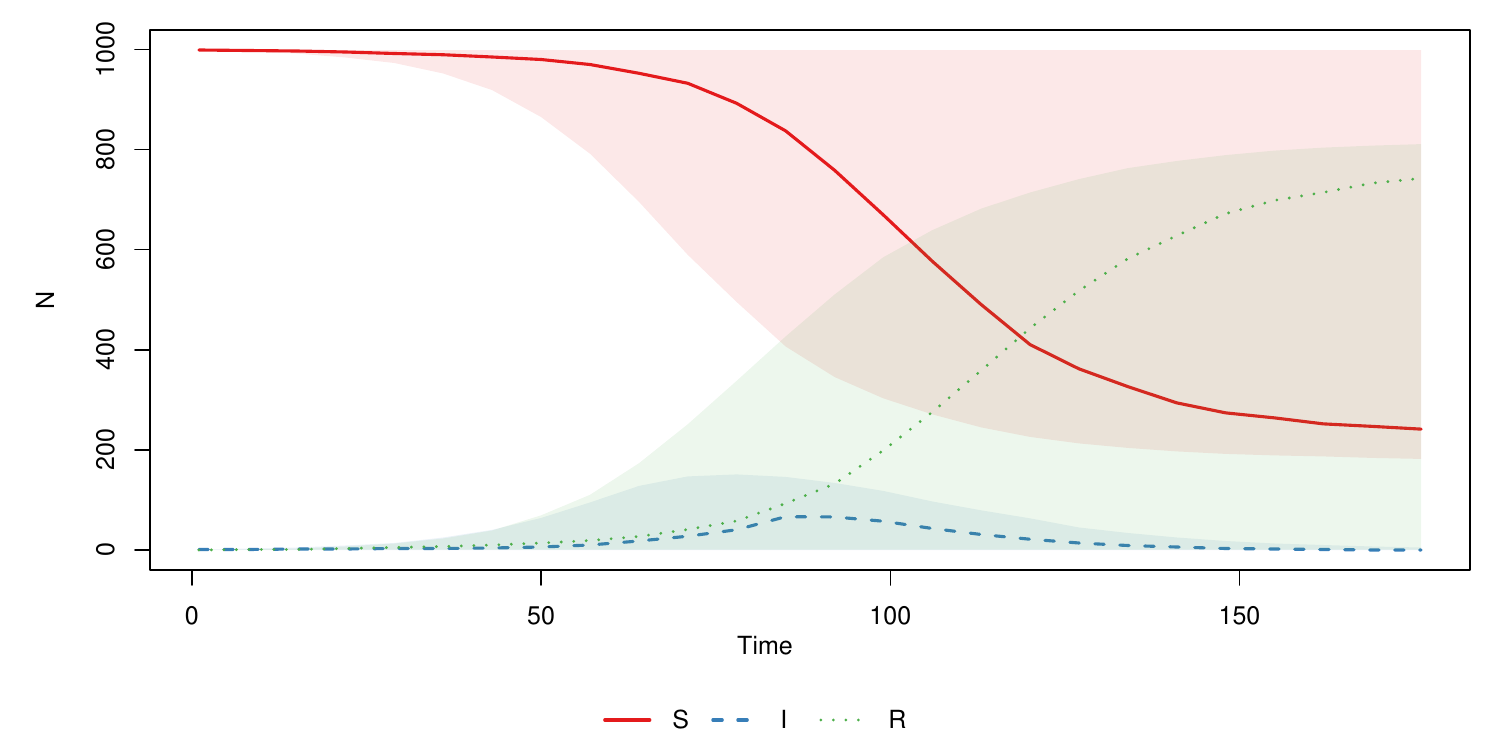}
  \includegraphics[width=0.49\linewidth]{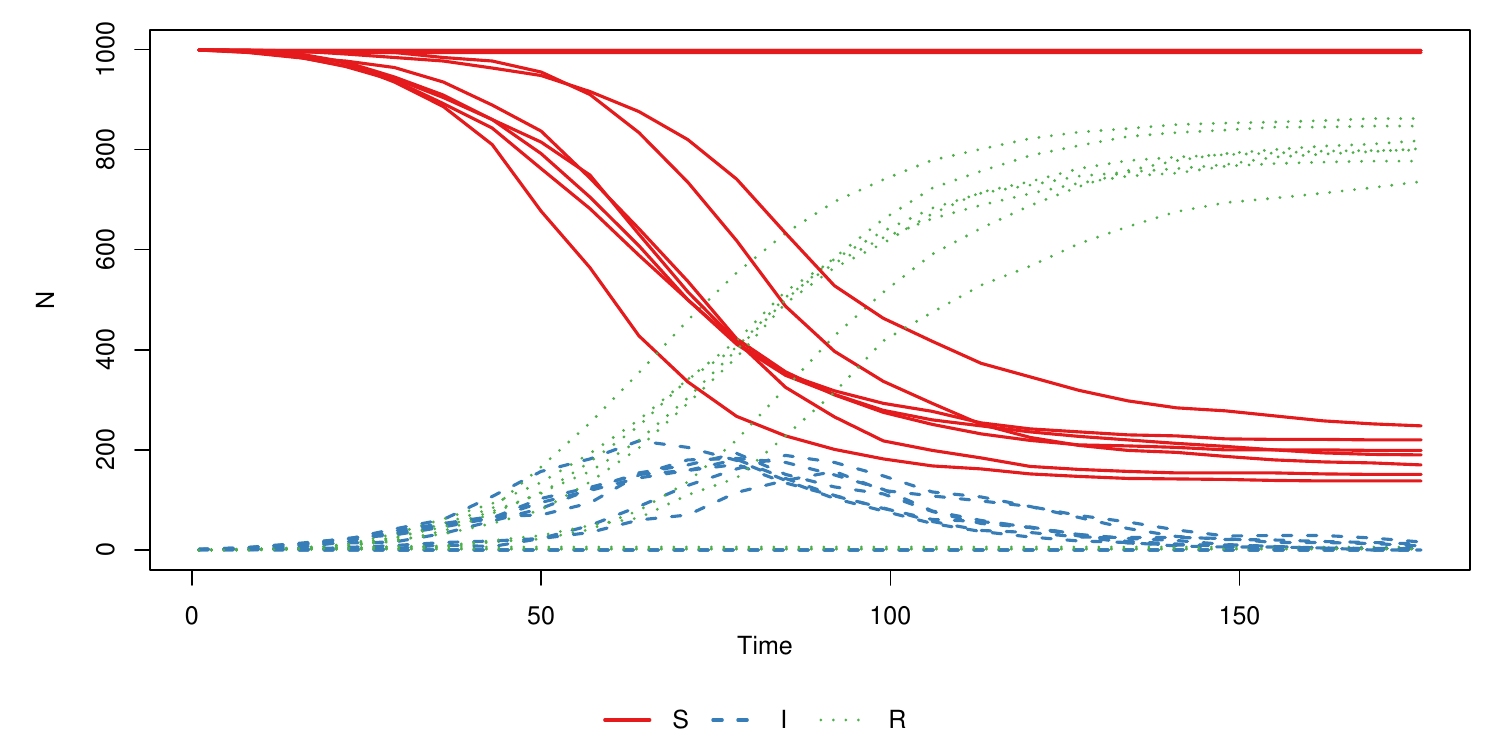}
  \end{center}
  \caption{Output from a stochastic \code{SIR} model in 1000 nodes
    starting with 999 susceptible, 1 infected and 0 recovered
    individuals in each node ($\beta = 0.16$, $\gamma = 0.077$).
    There are no between-node interactions.  Left: The default plot
    shows the median and inter-quartile range of the count in each
    compartment through time across all nodes.  Right: Realizations
    from a subset of 10 nodes.  \label{fig:SIR-model}}
\end{figure}

The \code{plot()} method of the \code{SimInf_model} class can be used
to visualize the simulated trajectory.  The default plot will display
the median count in each compartment across nodes as a colored line
together with the inter-quartile range using the same color, but with
transparency.  To display the outcome for individual nodes, specify
the subset of nodes to plot using the \code{node} parameter and set
\code{range = FALSE}.  In this example, an outbreak is likely to occur
in an infected node, but sometimes the infectious disease will become
extinct before it causes an epidemic, as shown in
Figure~\ref{fig:SIR-model}.

\begin{Schunk}
\begin{Sinput}
R> plot(result)
R> plot(result, node = 1:10, range = FALSE)
\end{Sinput}
\end{Schunk}

Most modeling and simulation studies require custom data analysis once
the simulation data has been generated.  To support this, \pkg{SimInf}
provides the \code{trajectory()} method to obtain a \code{data.frame}
with the number of individuals in each compartment at the time points
specified in \code{tspan}.  Below is an excerpt of the simulated data
from the first node that clearly shows there was an outbreak there.
To extract all data from every node, you only have to remove the node
argument.  Consult the help page for other \code{trajectory()}
parameter options.

\begin{Schunk}
\begin{Sinput}
R> head(trajectory(model = result, node = 1))
\end{Sinput}
\begin{Soutput}
  node time   S  I  R
1    1    1 999  1  0
2    1    8 998  1  1
3    1   15 991  8  1
4    1   22 973 21  6
5    1   29 935 42 23
6    1   36 886 61 53
\end{Soutput}
\end{Schunk}

\subsubsection[Specification of scheduled events in the SIR model]{%
  Specification of scheduled events in the \code{SIR} model}

In this example, we will continue to work with the predefined
\code{SIR} model to illustrate how demographic data can be
incorporated into a simulation.  In order for the numerical solver to
process a scheduled event, the compartments that are involved in the
event must be specified.  This is done by each event specifies one
column in the select matrix \code{E} using the \code{select} attribute
of the event.  The non-zero entries in the selected column in \code{E}
specify the involved compartments.  For the predefined \code{SIR}
model, \code{E} is defined as

{\small
\[
\mathbf{E} =
\bordermatrix{
    & 1 & 2 & 3 & 4 \cr
  S & 1 & 0 & 0 & 1 \cr
  I & 0 & 1 & 0 & 1 \cr
  R & 0 & 0 & 1 & 1 }
\]
}

This means that we can specify a scheduled event to operate on a
single compartment (S, I or R) as well as an event that involves all
three compartments.  When several compartments are involved in an
event, the individuals affected by the event will be sampled without
replacement from the specified compartments.  The numerical solver
performs an extensive error checking of the event before it is
processed.  And an error will be raised if the event is invalid, for
example, if the event tries to move more individuals than exists in
the specified compartments.

Consider we have 4 scheduled events to include in a simulation.  Below
is a \code{data.frame}, that contains the events.

\begin{Schunk}
\begin{Sinput}
R> events
\end{Sinput}
\begin{Soutput}
     event time node dest n proportion select shift
1    enter    2    3    0 5        0.0      1     0
2 extTrans    3    1    3 7        0.0      4     0
3     exit    4    2    0 0        0.2      4     0
4    enter    4    1    0 1        0.0      2     0
\end{Soutput}
\end{Schunk}

Interpret it as follows:

\begin{enumerate}
\itemsep0em
\item In time step 2 we add 5 susceptible individuals to node 3.
\item In time step 3 we sample 7 individuals without replacement among
  the S, I and R compartments in node 1 and move them to the
  corresponding compartments in node 3.
\item In time step 4 we sample 20\% of all individuals without
  replacement among the S, I and R compartments in node 2 and remove
  them from node 2.
\item In time step 4 we add 1 infected individual to node 1.
\end{enumerate}

Now, let us illustrate with a small example, consisting only of five
nodes, how scheduled events can alter the composition within nodes
during simulation.  Let us start with empty nodes and then create some
\textit{enter} events to add susceptible individuals to each node
during the first ten time-steps.

\begin{Schunk}
\begin{Sinput}
R> u0 <- data.frame(S = rep(0, 5), I = rep(0, 5), R = rep(0, 5))
R> add <- data.frame(event = "enter", time = rep(1:10, each = 5),
+    node = 1:5, dest = 0, n = 1:5, proportion = 0, select = 1, shift = 0)
\end{Sinput}
\end{Schunk}

We then create one \textit{enter} event to introduce an infected
individual to the 5$^{th}$ node at $t = 25$.

\begin{Schunk}
\begin{Sinput}
R> infect <- data.frame(event = "enter", time = 25, node = 5,
+    dest = 0, n = 1, proportion = 0, select = 2, shift = 0)
\end{Sinput}
\end{Schunk}

Additionally, we create some \textit{external transfer} events to form
interactions among the nodes with movements between $t = 35$ and $t =
45$.  Each shipment contains $n = 5$ individuals.

\begin{Schunk}
\begin{Sinput}
R> move <- data.frame(event = "extTrans", time = 35:45,
+    node = c(5, 5, 5, 5, 4, 4, 4, 3, 3, 2, 1), dest = c(4, 3, 3, 1, 3,
+    2, 1, 2, 1, 1, 2), n = 5, proportion = 0, select = 4, shift = 0)
\end{Sinput}
\end{Schunk}

Finally, we create \textit{exit} events to remove 20\% of the
individuals from each node at $t = 70$ and $t = 110$.

\begin{Schunk}
\begin{Sinput}
R> remove <- data.frame(event = "exit", time = c(70, 110),
+    node = rep(1:5, each = 2), dest = 0, n = 0, proportion = 0.2,
+    select = 4, shift = 0)
\end{Sinput}
\end{Schunk}

\begin{figure}
  \begin{center}
  \includegraphics[width=0.49\linewidth]{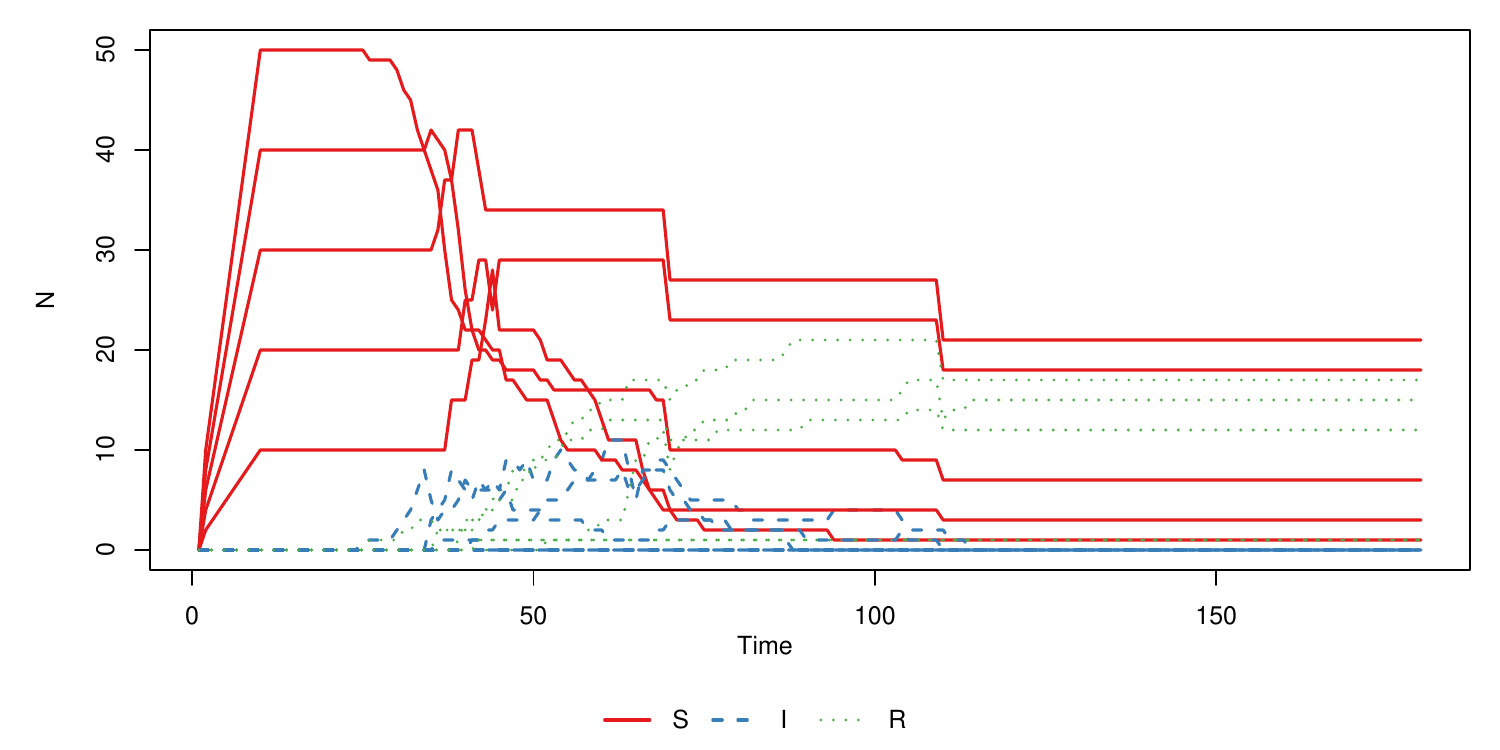}
  \includegraphics[width=0.49\linewidth]{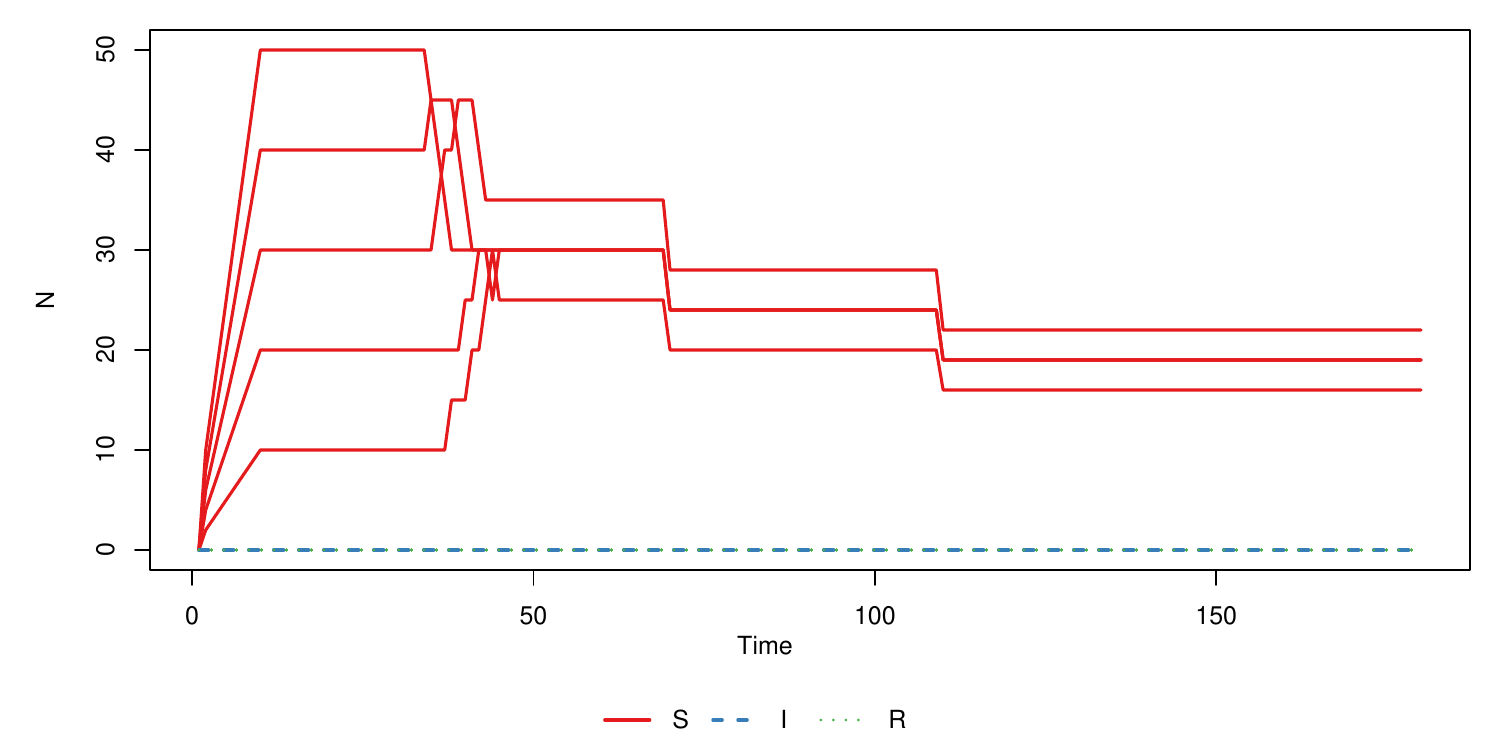}
  \end{center}
  \caption{An \code{SIR} model in five nodes with scheduled events.
    Left: One realization where susceptible individuals were added
    during the first ten time-steps.  Then, one infected individual
    was introduced at $t = 25$.  Further, some movements occurred
    between $t = 35$ and $t = 45$.  Finally, at $t = 70$ and $t =
    110$, twenty percent of the individuals were removed.  Right: For
    comparison, the deterministic dynamics when not introducing an
    infected individual.}
  \label{fig:SIR-events}
\end{figure}

Figure~\ref{fig:SIR-events} shows one realization of a model
incorporating the events.  Here we observe two transmission processes
on different scales.  First, the stochastic transmission process
within each node.  Secondly, the between-node transmission due to
movements.  Also stochastic, because of the sampling process that
select susceptible, infected or recovered individuals to move between
the nodes.

\begin{Schunk}
\begin{Sinput}
R> events = rbind(add, infect, move, remove)
R> model <- SIR(u0 = u0, tspan = 1:180, events = events, beta = 0.16,
+    gamma = 0.077)
R> set.seed(3)
R> result <- run(model, threads = 1)
R> plot(result, node = 1:5, range = FALSE)
\end{Sinput}
\end{Schunk}

We can use \code{replicate} (or similar) to generate many realizations
from a \code{SimInf_model} object, together with some custom analysis
of each trajectory.  Here, we find that infection spread from the
5$^{th}$ node in about half of \code{n = 1000} trajectories.

\begin{Schunk}
\begin{Sinput}
R> set.seed(123)
R> mean(replicate(n = 1000, {
+    nI <- trajectory(run(model = model, threads = 1), node = 1:4)$I
+    sum(nI) > 0
+  }))
\end{Sinput}
\begin{Soutput}
[1] 0.486
\end{Soutput}
\end{Schunk}

\subsubsection[C code for the SIR model]{%
  \proglang{C} code for the \code{SIR} model}

The \proglang{C} code for the \code{SIR} model is defined in the
source file \texttt{'src/models/SIR.c'}.  This file contains the
\code{SIR_run} function to initialize the core solver
(Listing~\ref{lst:SIRrun} in Appendix~\ref{sec:C-code-SIR}), the
transition rate functions (Listing~\ref{lst:trSIR} in
Appendix~\ref{sec:C-code-SIR}) and the post time step function
(Listing~\ref{lst:ptsSIR} in Appendix~\ref{sec:C-code-SIR}).

\subsection[A second example: The SISe_sp model]{%
  A second example: The \code{SISe\_sp} model}
\label{sec:example-SISe_sp}

Here we will illustrate the use of local data (\code{ldata}) and
continuous state variables (\code{V}) to formulate a more complex
model with variables obeying ODEs.  Moreover, we will introduce the
\code{prevalence()} method, another important function for
post-processing trajectories.  Let us consider VTEC in cattle for this
example. Briefly, a VTEC infection in cattle can be formulated as a
susceptible-infected-susceptible (SIS) compartment model.  However,
previous modeling has shown that it is important to consider within
and between-farm transmission via the environment \citep{Ayscue2009,
  Zhang2010}, ambient temperature \citep{Gautam2011}, herd size and
between-farm spread from livestock movements \citep{Zhang2010}.
Therefore, let us use the predefined \code{SISe\_sp} model.  It
contains an environmental compartment to model shedding of a pathogen
to the environment.  Moreover, it also includes a spatial coupling of
the environmental contamination among proximal nodes to capture
between-node spread unrelated to moving infected individuals.
Consequently, the model has two state transitions,
\begin{align}
\label{eq:vtectrans}
\begin{array}{rcl}
  S_i & \xrightarrow{\upsilon \varphi_i} & I_i, \\
  I_i & \xrightarrow{\gamma} & S_i,
  \end{array}
\end{align}
where the transition rate per unit of time from susceptible to
infected is proportional to the concentration of the environmental
contamination $\varphi_i(t)$ in node $i$.  Moreover, the transition
rate from infected to susceptible is the recovery rate $\gamma$,
measured per individual and per unit of time.  Finally, the
environmental infectious pressure is evolved by
\begin{equation}
  \label{eq:envInfPressure-local-spread}
  \frac{d \varphi_i(t)}{dt}= \frac{\alpha I_i(t)}{N_i(t)} +
  \sum_k{\frac{\varphi_k(t) N_k(t) - \varphi_i(t) N_i(t)}{N_i(t)}
    \cdot \frac{D}{d_{ik}}} - \beta(t) \varphi_i(t),
\end{equation}
where $\alpha$ is the average shedding rate of the pathogen to the
environment per infected individual and $N_i = S_i + I_i$ the size of
node $i$.  Next comes the spatial coupling among proximal nodes, where
$D$ is the rate of the local spread and $d_{ik}$ the distance between
holdings $i$ and $k$.  The seasonal decay and removal of the pathogen
is captured by $\beta(t)$.  The environmental infectious pressure
$\varphi_i(t)$ in each node is evolved in the post-time-step function
by the Euler forward method (see the file 'src/models/SISe\_sp.c' for
the \proglang{C} code).  The value of $\varphi_i(t)$ is saved to the
\code{V} matrix at the time-points specified by \code{tspan}.

Let us use a synthetic dataset of 1600 farms located within a 50
square kilometer region.  Load the data with the following commands

\begin{Schunk}
\begin{Sinput}
R> data("nodes", package = "SimInf")
R> u0 <- u0_SISe()
R> events <- events_SISe()
\end{Sinput}
\end{Schunk}

where the location of each farm is in \code{nodes}, \code{u0} defines
the initial cattle population and \code{events} contains four years
(1460 days) of scheduled events data (births, deaths and movements).
Moreover, let us define proximal neighbors as neighbors within 2500m
and use the utility function \code{distance\_matrix()} to estimate the
distance between farms within that cutoff.

\begin{Schunk}
\begin{Sinput}
R> d_ik <- distance_matrix(x = nodes$x, y = nodes$y, cutoff = 2500)
\end{Sinput}
\end{Schunk}

Let us assume that 10\% of the farms have 5\% infected cattle at the
beginning of the simulation.

\begin{Schunk}
\begin{Sinput}
R> set.seed(123)
R> i <- sample(x = 1:1600, size = 160)
R> u0$I[i] <- as.integer(u0$S[i] * 0.05)
R> u0$S[i] <- u0$S[i] - u0$I[i]
\end{Sinput}
\end{Schunk}

The \code{SISe\_sp} model contains parameters at a global and local
scale.  Here, the parameter values were chosen such that the
proportion of infected nodes in a trajectory is about 10\% and
displays a seasonal pattern.  The global parameters are: the spatial
\code{coupling = 0.2} ($D$ in
Equation~\eqref{eq:envInfPressure-local-spread}), the shedding rate
\code{alpha = 1}, the recovery rate \code{gamma = 0.1} and the
indirect transmission rate \code{upsilon = 0.012}.  Moreover, the
global parameter $\beta(t)$ captures decay of the pathogen in four
seasons: \code{beta\_t1 = 0.1}, \code{beta\_t2 = 0.12}, \code{beta\_t3
  = 0.12} and \code{beta\_t4 = 0.1}.  However, the duration of each
season is local to a node and is specified as the day of the year each
season ends.  Here, for simplicity, we let \code{end\_t1 = 91},
\code{end\_t2 = 182}, \code{end\_t3 = 273} and \code{end\_t4 = 365} in
all nodes.  Furthermore, the distances between nodes are local data
extracted from \code{distance = d\_ik}.  Finally, we let \code{phi =
  0} at the beginning of the simulation (becomes \code{v0} in the
model object).

\begin{Schunk}
\begin{Sinput}
R> model <- SISe_sp(u0 = u0, tspan = 1:1460, events = events, phi = 0,
+    upsilon = 0.012, gamma = 0.1, alpha = 1, beta_t1 = 0.10,
+    beta_t2 = 0.12, beta_t3 = 0.12, beta_t4 = 0.10, end_t1 = 91,
+    end_t2 = 182, end_t3 = 273, end_t4 = 365, distance = d_ik,
+    coupling = 0.2)
\end{Sinput}
\end{Schunk}

Let us use the \code{prevalence()} method to explore the proportion of
infected nodes through time.  It takes a model object and a formula
specification, where the left hand side of the formula specifies the
compartments representing cases i.e., have an attribute or a disease.
The right hand side of the formula specifies the compartments at risk.
Here, we are interested in the proportion of nodes with at least one
infected individual, therefore, we let \code{formula = I ~ S + I} and
specify \code{type = "nop"}.  Consult the help page for other
\code{prevalence()} parameter options.

\begin{Schunk}
\begin{Sinput}
R> plot(NULL, xlim = c(0, 1500), ylim = c(0, 0.18),
+    ylab = "Prevalance", xlab = "Time")
R> set.seed(123)
R> replicate(5, {
+      result <- run(model = model, threads = 1)
+      p <- prevalence(model = result, formula = I ~ S + I, type = "nop")
+      lines(p)
+    })
\end{Sinput}
\end{Schunk}

Assume there exists some sort of treatment by which the
\code{coupling} can be reduced by 50\%. Is that sufficient for
controlling the disease?  We can use the function \code{gdata()} to
change the global \code{coupling} parameter and then run some more
trajectories.

\begin{Schunk}
\begin{Sinput}
R> gdata(model, "coupling") <- 0.1
R> replicate(5, {
+      result <- run(model = model, threads = 1)
+      p <- prevalence(model = result, formula = I ~ S + I, type = "nop")
+      lines(p, col = "blue", lty = 2)
+    })
\end{Sinput}
\end{Schunk}

\begin{figure}
  \begin{center}
  \includegraphics{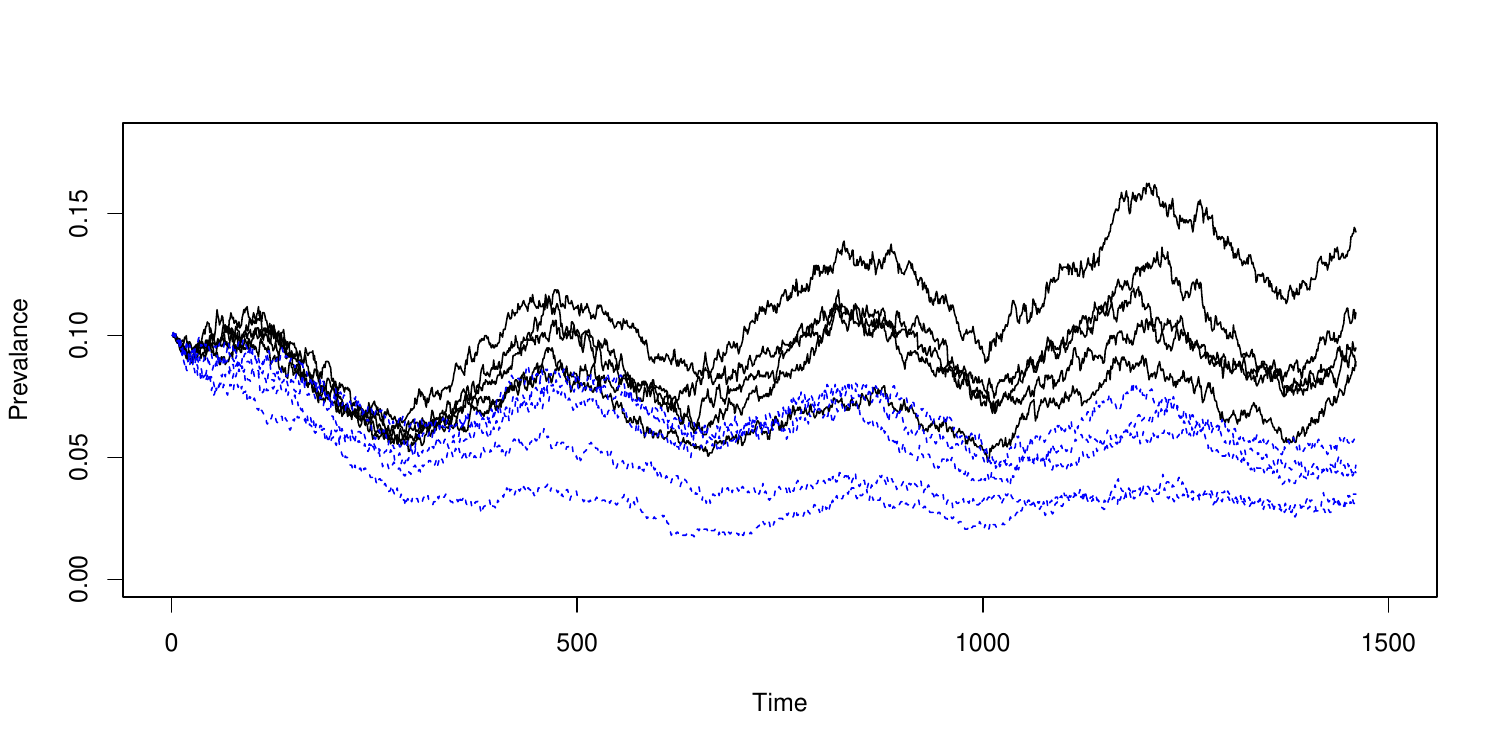}
  \end{center}
  \caption{Exploring options for reducing disease spread. Reference
    trajectories showing the proportion infected nodes (black solid
    lines).  The proportion infected nodes after reducing the spatial
    coupling with 50\% (blue dotted lines).}
  \label{fig:SISe_sp}
\end{figure}

The results in Figure~\ref{fig:SISe_sp} indicate that reducing the
spatial coupling $D$ with 50\% is not sufficient to eradicate the
infection from this synthetic cattle population.  Nevertheless this
short example serves as a template for using the \pkg{SimInf}
computational framework when implementing large scale data-driven
disease spread models and exploring options for control.


\section[Extending SimInf: New models]{Extending \pkg{SimInf}: New models}
\label{sec:extend}

One of the design goals of \pkg{SimInf} was to make it extendable.
The current design supports two ways to extend \pkg{SimInf} with new
models, and this section describes the relevant steps to implement a
new model.  Since extending \pkg{SimInf} requires that \proglang{C}
code can be compiled, you will first need to install a compiler.  To
read more about interfacing compiled code from \proglang{R} and
creating \proglang{R} add-on packages, the \texttt{'Writing R
  extensions'}
(\url{https://cran.r-project.org/doc/manuals/r-release/R-exts.html})
manual is the official guide and describes the process in detail.
Another useful resource is the \texttt{'R packages'} book by
\cite{Hadley2015} (\url{http://r-pkgs.had.co.nz/}).

\subsection{Using the model parser to define a new model}
\label{sec:mparse}

The simplest way to define a new model for \pkg{SimInf} is to use the
model parser method \code{mparse}.  It takes a character vector of
transitions in the form of \code{"X -> propensity -> Y"} and generates
the \proglang{C} and \proglang{R} code for the model.  The left hand
side of the first '\code{->}'-sign is the initial state, the right
hand side of the last '\code{->}'-sign is the final state, and the
propensity is written between the '\code{->}'-signs.  The special
symbol '\code{@}' is reserved for the empty set $\emptyset$.  We
suggest to first draw a schematic representation of the model that
includes all compartments and arrows for all state transitions.

\subsubsection[Introductory examples of using mparse]{%
  Introductory examples of using \code{mparse}}
\label{sec:mparse-introductory-examples}

In a first example we will consider the SIR model in a closed
population i.e., no births or deaths.  If we let \code{b} denote the
transmission rate and \code{g} the recovery rate, the model can be
described as,

\begin{Schunk}
\begin{Sinput}
R> transitions <- c("S -> b*S*I/(S+I+R) -> I", "I -> g*I -> R")
R> compartments <- c("S", "I", "R")
\end{Sinput}
\end{Schunk}

We can now use the \code{transitions} and \code{compartments}
variables, together with the constants \code{b} and \code{g} to build
an object of class \code{'SimInf\_model'} via a call to \code{mparse}.
It also needs to be initialized with the initial condition \code{u0}
and \code{tspan}.

\begin{Schunk}
\begin{Sinput}
R> n <- 1000
R> u0 <- data.frame(S = rep(99, n), I = rep(5, n), R = rep(0, n))
R> model <- mparse(transitions = transitions, compartments = compartments,
+    gdata = c(b = 0.16, g = 0.077), u0 = u0, tspan = 1:180)
\end{Sinput}
\end{Schunk}

As in earlier examples, the \code{model} object can now be used to
simulate data and plot the results.  Internally, the \proglang{C} code
that was generated by \code{mparse} is written to a temporary file
when the \code{run} method is called.  If the temporary file is
compiled successfully, the resulting DLL is dynamically loaded and
used to run one trajectory of the model.  Once the simulator
completes, the DLL is unloaded and the temporary files are removed.

\begin{Schunk}
\begin{Sinput}
R> set.seed(123)
R> result <- run(model = model, threads = 1)
R> plot(result)
\end{Sinput}
\end{Schunk}

\begin{figure}
  \begin{center}
\includegraphics{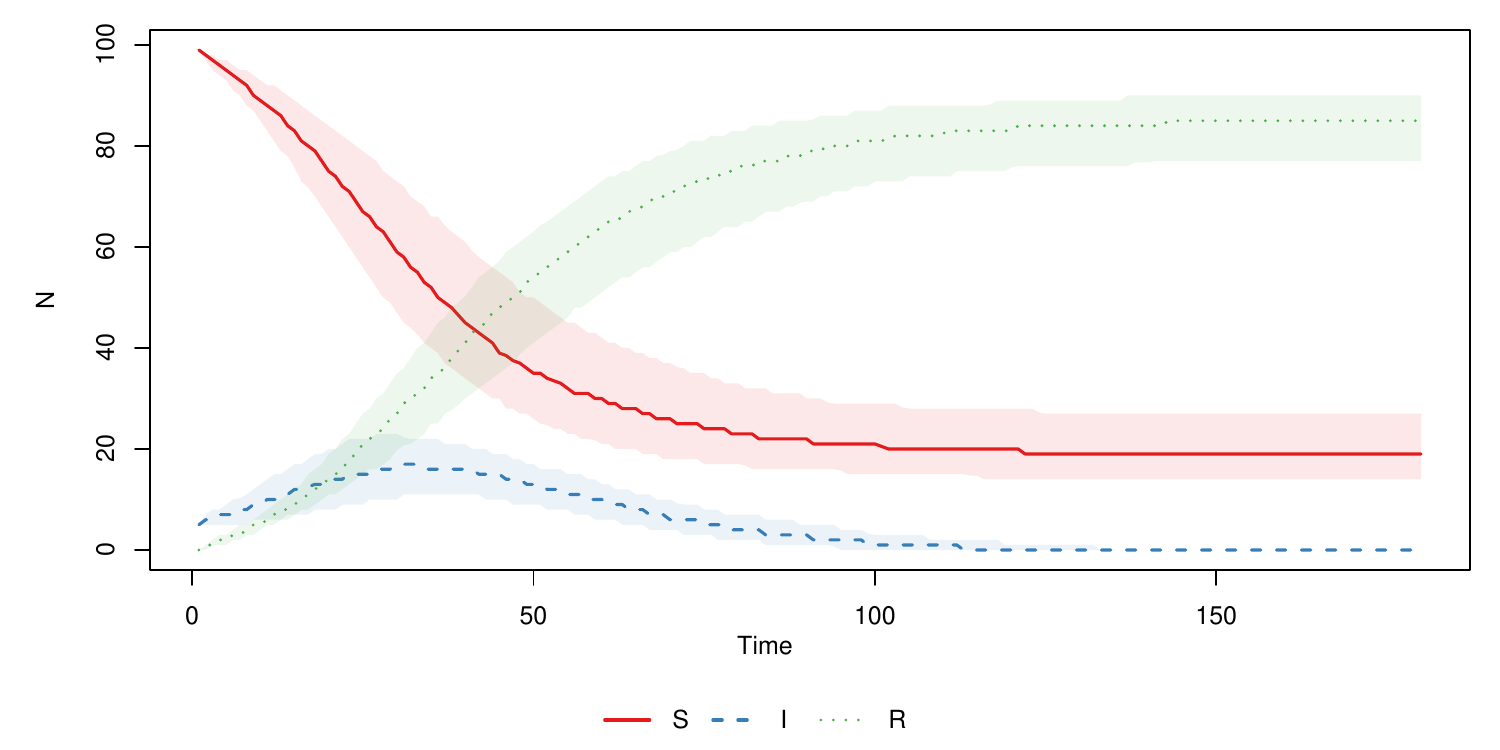}
  \caption{Showing the median and inter-quartile range from 1000
    realizations of an \code{mparse} \code{SIR} model ($\beta = 0.16$,
    $\gamma = 0.077$), starting with 99 susceptible, 5 infected, and 0
    recovered individuals. \label{fig:SIR-mparse-proportion}}
  \end{center}
\end{figure}

The flexibility of the \code{mparse} approach allows for quick
prototyping of new models or features.  Let us elaborate on the
previous example and explore the incidence cases per day.  This can
easily be done by adding a new compartment \code{'Icum'} whose sole
purpose is to keep track of how many individuals who become infected
over time.  The right hand side \code{'I + Icum'} of the transition
\code{'S -> b*S*I/(S+I+R) -> I + Icum'}, means that both \code{'I'}
and \code{'Icum'} are incremented by one each time the transition
happens.

\begin{Schunk}
\begin{Sinput}
R> transitions <- c("S -> b*S*I/(S+I+R) -> I + Icum", "I -> g*I -> R")
R> compartments <- c("S", "I", "Icum", "R")
\end{Sinput}
\end{Schunk}

Since there are no between-node movements in this example, the
stochastic process in one node does not affect any other nodes in the
model.  It is therefore straightforward to run many realizations of
this model, simply by replicating a node in the initial condition
\code{u0}, for example, $n = 1000$ times.

\begin{Schunk}
\begin{Sinput}
R> n <- 1000
R> u0 <- data.frame(S = rep(99, n), I = rep(1, n), Icum = rep(0, n),
+    R = rep(0, n))
R> model <- mparse(transitions = transitions, compartments = compartments,
+    gdata = c(b = 0.16, g = 0.077), u0 = u0, tspan = 1:150)
R> set.seed(123)
R> result <- run(model = model, threads = 1)
\end{Sinput}
\end{Schunk}

Let us post-process the simulated trajectory to compare the incidence
cases in the first node with the average incidence cases among all
realizations by extracting the trajectory data and calculate
successive differences of \code{'Icum'} at each time-point.

\begin{Schunk}
\begin{Sinput}
R> traj <- trajectory(model = result, compartments = "Icum")
R> cases <- stepfun(result@tspan[-1], diff(c(0, traj$Icum[traj$node == 1])))
R> avg_cases <- c(0, diff(by(traj, traj$time, function(x) sum(x$Icum))) / n)
\end{Sinput}
\end{Schunk}

\begin{figure}
  \begin{center}
\includegraphics{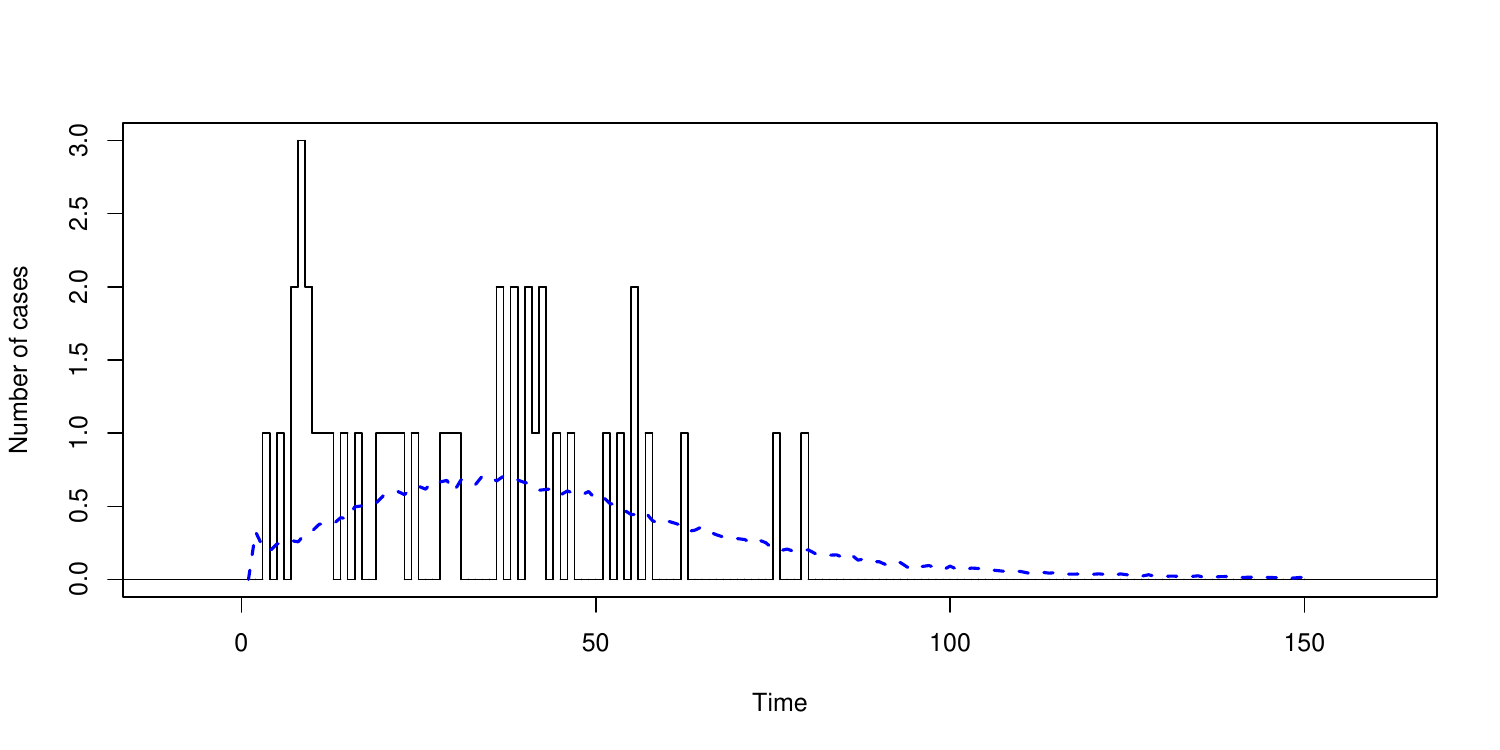}
  \end{center}
  \caption{(black solid line) One realization of an epidemic curve
    displaying the number of incident cases per day in a node when
    simulating 150 days of an \code{mparse} \code{SIR} model ($\beta =
    0.16, \gamma = 0.077$), starting with 99 susceptible, 1 infected
    and 0 recovered individuals.  (blue dashed line) Average number of
    incident cases per day from 1000 realizations of the
    model.  \label{fig:SIR-mparse-incidence}}
\end{figure}

Finally, plot the result as an epidemic curve
(Figure~\ref{fig:SIR-mparse-incidence}).  In this example, the number
of incident cases in the first node exceeds what is expected on
average.

\begin{Schunk}
\begin{Sinput}
R> plot(cases, main = "", xlab = "Time", ylab = "Number of cases",
+    do.points = FALSE)
R> lines(avg_cases, col = "blue", lwd = 2, lty = 2)
\end{Sinput}
\end{Schunk}

\subsubsection[Incorporate scheduled events in an mparse model]{%
  Incorporate scheduled events in an \code{mparse} model}
\label{sec:mparse-scheduled-events}

To illustrate how models generated using \code{mparse} can incorporate
scheduled events, consider an epidemic in a population consisting of
1600 nodes, for example, cattle herds, that are connected to each
other by livestock movements.  Assume an outbreak is detected on day
twenty-one after introduction of an infection in one node and that we
wish to explore how vaccination could limit the outbreak, if resources
for vaccination can handle 50 herds per day and 80\% of the animals in
each herd.  Let us add a new compartment $V$ to the model to represent
vaccinated individuals, so that the model now contains the $\{S, I,
I_{cum}, R, V \}$ compartments.  As before, let \code{b} denote the
transmission rate and \code{g} the recovery rate.

\begin{Schunk}
\begin{Sinput}
R> transitions <- c("S -> b*S*I/(S+I+R+V) -> I + Icum", "I -> g*I -> R")
R> compartments <- c("S", "I", "Icum", "R", "V")
\end{Sinput}
\end{Schunk}

Load the example data for an \code{SIR} model in a population of 1600
nodes (cattle herds) with its associated scheduled events: births,
deaths, and livestock movements.  Moreover, let $I_{cum} = 0$ and $R =
0$.

\begin{Schunk}
\begin{Sinput}
R> u0 <- u0_SIR()
R> u0$Icum <- 0
R> u0$V <- 0
R> events <- events_SIR()
\end{Sinput}
\end{Schunk}

Now generate vaccination events i.e., \textit{internal transfer}
events.  Use \code{select = 3} and \code{shift = 1} to move animals
from the susceptible, infectious and recovered compartments to the
vaccinated compartment, see the definitions of $E$ and $N$ below.  Let
us start the vaccinations in nodes 1--50 on day twenty-one, and
continue until all herds are vaccinated on day fifty-two.  Moreover,
use \code{proportion = 0.8} to vaccinate 80\% of the animals in each
herd.  We assume, for the sake of simplicity, that vaccinated
individuals become immune and non-infectious immediately.

\begin{Schunk}
\begin{Sinput}
R> vaccination <- data.frame(event = "intTrans", time = rep(21:52, each = 50),
+    node = 1:1600, dest = 0, n = 0, proportion = 0.8, select = 3,
+    shift = 1)
\end{Sinput}
\end{Schunk}

To simulate from this model, we have to define the select matrix $E$
to handle which compartments to sample from when processing a
scheduled event.  Let the first column in $E$ handle \textit{enter}
events (births); add newborn animals to the susceptible compartment
$S$.  The second column is for \textit{exit} events (deaths) and
\textit{external transfer} events (livestock movements); sample
animals from the $S$, $I$, $R$ and $V$ compartments.  Finally, the
third column is for \textit{internal transfer} events (vaccination);
sample individuals from the $S$, $I$ and $R$ compartments.  We must
also define the shift matrix $N$ to process \textit{internal transfer}
events (vaccination); move sampled animals from the $S$ compartment
four steps forward to the $V$ compartment.  Similarly, move sampled
animals from the $I$ compartment three steps forward to the $V$
compartment, and finally, move sampled individuals from the $R$
compartment one step forward.

{\small
\[
\mathbf{E} =
\bordermatrix{
         & 1 & 2 & 3 \cr
  S      & 1 & 1 & 1 \cr
  I      & 0 & 1 & 1 \cr
  I_{cum} & 0 & 0 & 0 \cr
  R      & 0 & 1 & 1 \cr
  V      & 0 & 1 & 0} \qquad
\mathbf{N} =
\bordermatrix{
         & 1 \cr
  S      & 4 \cr
  I      & 3 \cr
  I_{cum} & 0 \cr
  R      & 1 \cr
  V      & 0}
\]
}

\begin{Schunk}
\begin{Sinput}
R> E <- matrix(c(1, 0, 0, 0, 0, 1, 1, 0, 1, 1, 1, 1, 0, 1, 0), nrow = 5,
+    ncol = 3, dimnames = list(c("S", "I", "Icum", "R", "V"),
+    c("1", "2", "3")))
R> N <- matrix(c(4, 3, 0, 1, 0), nrow = 5, ncol = 1,
+    dimnames = list(c("S", "I", "Icum", "R", "V"), "1"))
\end{Sinput}
\end{Schunk}

Additionally, we have to redefine the \code{select} column in the
events, since the data is from the predefined \code{SIR} model with
another $E$ matrix.  Let us change \code{select} for movements.
\begin{Schunk}
\begin{Sinput}
R> events$select[events$select == 4] <- 2
\end{Sinput}
\end{Schunk}

Now let us create an \code{epicurve} function to estimate the average
number of new cases per day from \code{n = 1000} realizations in a
\code{for}-loop and simulate one trajectory at a time.  To clear
infection that was introduced in the previous trajectory, animals are
first moved to the susceptible compartment.  Then, one infected
individual is introduced into a randomly sampled node from the
population.  Note that we use the 'L' suffix to create an integer
value rather than a numeric value.  Run the model and accumulate
\code{Icum}.  For efficiency, use \code{as.is = TRUE}, the internal
matrix format, to extract \code{Icum} in every node at each time-point
in \code{tspan}.

\begin{Schunk}
\begin{Sinput}
R> epicurve <- function(model, n = 1000) {
+    Icum <- numeric(length(model@tspan))
+    for (i in seq_len(n)) {
+      model@u0["S", ] <- model@u0["S", ] + model@u0["I", ]
+      model@u0["I", ] <- 0L
+      j <- sample(seq_len(Nn(model)), 1)
+      model@u0["I", j] <- 1L
+      model@u0["S", j] <- model@u0["S", j] - 1L
+      result <- run(model = model)
+      traj <- trajectory(model = result, compartments = "Icum",
+        as.is = TRUE)
+      Icum <- Icum + colSums(traj)
+    }
+    stepfun(model@tspan[-1], diff(c(0, Icum / n)))
+  }
\end{Sinput}
\end{Schunk}

Generate an epicurve with the average number of cases per day for the
first three hundred days of the epidemic without vaccination.

\begin{Schunk}
\begin{Sinput}
R> model_no_vac <- mparse(transitions = transitions,
+    compartments = compartments, gdata = c(b = 0.16, g = 0.077),
+    u0 = u0, tspan = 1:300, events = events, E = E, N = N)
R> cases_no_vac <- epicurve(model_no_vac)
\end{Sinput}
\end{Schunk}

Similarly, generate an epicurve after incorporating the
\code{vaccination} events.

\begin{Schunk}
\begin{Sinput}
R> model_vac <- mparse(transitions = transitions,
+    compartments = compartments, gdata = c(b = 0.16, g = 0.077),
+    u0 = u0, tspan = 1:300, events = rbind(events, vaccination),
+    E = E, N = N)
R> cases_vac <- epicurve(model_vac)
\end{Sinput}
\end{Schunk}

\begin{figure}
  \begin{center}
    \includegraphics{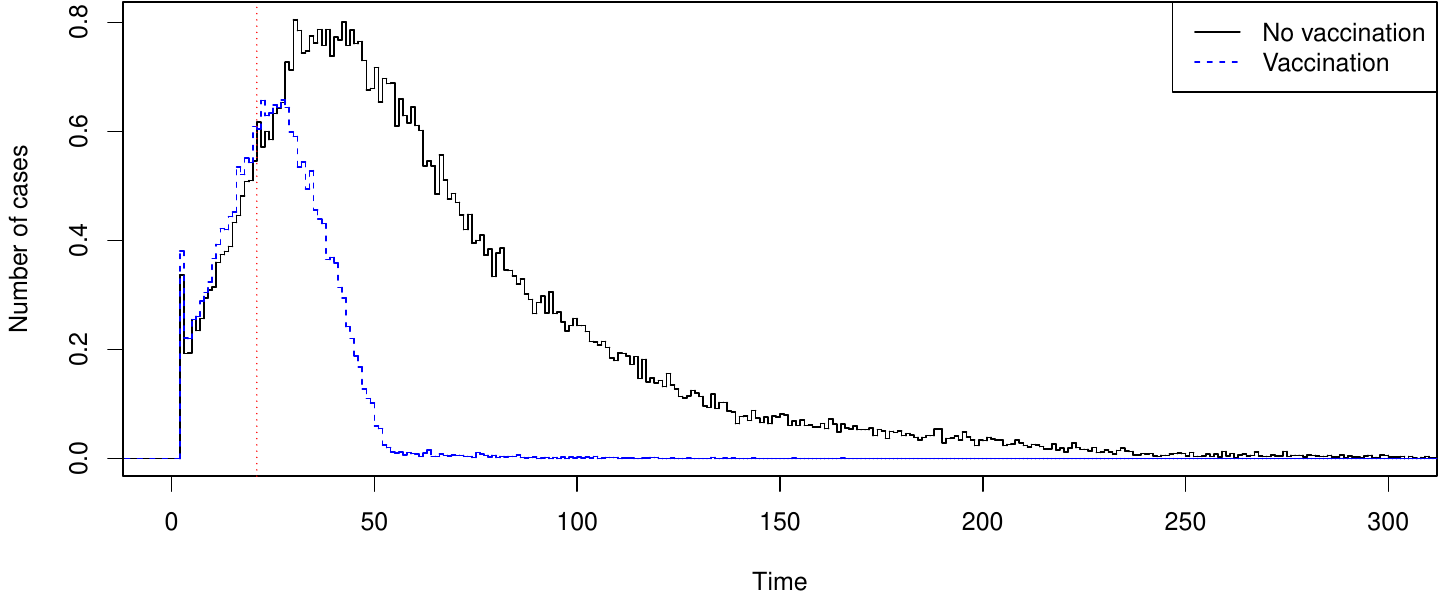}
  \end{center}
  \caption{Comparison beween the number of cases per day of an
    outbreak in an unvaccinated population of cattle herds (black
    solid line) and after vaccination of animals (blue dashed line).
    The vertical line (dotted) indicates when vaccination was
    initiated.  \label{fig:mparse-epicurve}}
\end{figure}

As expected, the number of cases decrease rapidly after vaccination,
while the outbreak is ongoing for a longer time in the unvaccinated
population (Figure~\ref{fig:mparse-epicurve}).

\begin{Schunk}
\begin{Sinput}
R> plot(cases_no_vac, main = "", xlim = c(0, 300), xlab = "Time",
+    ylab = "Number of cases", do.points = FALSE)
R> lines(cases_vac, col = "blue", do.points = FALSE, lty = 2)
R> abline(v = 21, col = "red", lty = 3)
R> legend("topright", c("No vaccination", "Vaccination"),
+    col = c("black", "blue"), lty = 1:2)
\end{Sinput}
\end{Schunk}

\subsection[Use the SimInf framework from another package]{%
  Use the \pkg{SimInf} framework from another package}
\label{sec:linking-to}

Another possibility is to extend \pkg{SimInf} by creating an
\proglang{R} add-on package that uses \pkg{SimInf} by linking to its
core solver native routine.  To facilitate this, the \pkg{SimInf}
package includes the \code{package_skeleton} method to automate some
of the setup for a new source package.  It creates directories, saves
\proglang{R} and \proglang{C} code files to appropriate places, and
creates skeleton help files.

Even if \pkg{SimInf} was designed to study the dynamics of infectious
diseases, it is not limited to that use case but can be used to study
the dynamics of other systems.  Consider we wish to create a new
add-on package \pkg{PredatorPrey} based on the Rosenzweig-MacArthur
predator-prey model demonstrated in the \pkg{GillespieSSA} package
\citep{Rosenzweig1963, Pineda-Krch2008}.  The model has a
density-dependent growth in the prey and and a nonlinear Type-2
functional response in the predator \citep{Rosenzweig1963}.  Let $R$
and $F$ denote the number of prey and predators, respectively.  The
model consists of five transitions
(Equation~\eqref{eq:predator-prey}): \textit{i)} prey birth,
\textit{ii)} prey death due to non-predatory events, \textit{iii)}
prey death due to predation, \textit{iv)} predator birth, and
\textit{v)} predator death

\begin{align}
  \label{eq:predator-prey}
  \left.
     \begin{array}{rcl}
      \emptyset & \xrightarrow{b_R \cdot R} & R \\
      R & \xrightarrow{(d_R+(b_R-d_R) \cdot R/K) \cdot R} & \emptyset \\
      R & \xrightarrow{\alpha/(1+w \cdot R) \cdot R \cdot F} & \emptyset \\
      \emptyset & \xrightarrow{b_F \cdot \alpha/(1+w \cdot R) \cdot R \cdot F} & F \\
      F & \xrightarrow{d_F \cdot F} & \emptyset \\
    \end{array}
  \right\},
\end{align}

where $b_R$, $d_R$, $b_F$, and $d_F$ are the per capita birth and
death rate of the prey and predator, respectively.  Moreover, $K$ is
the carrying capacity of the prey, $\alpha$ is the predation
efficiency, and $w$ is the degree of predator saturation
\citep{Pineda-Krch2008}.  Using parameter values from
\cite{Pineda-Krch2008}, we define the model as

\begin{Schunk}
\begin{Sinput}
R> transitions <- c("@ -> bR*R -> R", "R -> (dR+(bR-dR)*R/K)*R -> @",
+    "R -> alpha/(1+w*R)*R*F -> @", "@ -> bF*alpha/(1+w*R)*R*F -> F",
+    "F -> dF*F -> @")
R> compartments <- c("R", "F")
R> parameters <- c(bR = 2, bF = 2, dR = 1, K = 1000, alpha = 0.007,
+    w = 0.0035, dF = 2)
\end{Sinput}
\end{Schunk}

Assume the initial population consists of $R = 1000$ prey and $F =
100$ predators and we are interested in simulating $n = 1000$
replicates over 100 days.  Since there are no between-node movements
in this example, we can generate replicates simply by starting with
$n$ identical nodes.

\begin{Schunk}
\begin{Sinput}
R> n <- 1000
R> u0 = data.frame(R = rep(1000, n), F = rep(100, n))
R> model <- mparse(transitions = transitions, compartments = compartments,
+    gdata = parameters, u0 = u0, tspan = 1:100)
\end{Sinput}
\end{Schunk}

Now instead of running the model to generate data, let us use it to
create an \proglang{R} add-on package.

\begin{Schunk}
\begin{Sinput}
R> path <- tempdir()
R> package_skeleton(model = model, name = "PredatorPrey", path = path)
\end{Sinput}
\end{Schunk}

Where the first argument is the \code{SimInf\_model} object generated
by the \code{mparse} method, the second argument is the name of the
package to create a skeleton for and the third argument is the path to
the new package. Note that a temporary directory is used here for
illustration of the functionality.  We refer to the \pkg{SimInf}
documentation for other arguments that can be supplied to the
\code{package_skeleton} method.  The created \proglang{R} file
(\texttt{'R/models.R'}) defines the \code{S4} class
\code{PredatorPrey} that contains the \code{SimInf\_model} and a
generating function to create a new object of the \code{PredatorPrey}
model.  The generating function is a template that might need to be
extended to meet the specific requirements for the model.

The \proglang{C} file (\texttt{'src/models.c'}) defines one function
for each state transition, the post time step function and the model
specific run function. The file is automatically compiled when
installing the package.  The header file \code{"SimInf.h"} contains
the declarations for these functions and must be included.  The
\code{SimInf_model_run} function is the interface from \proglang{R} to
the core solver in \proglang{C} and list all function pointers to the
transition rate functions in a vector in the order the state
transitions appear in the dependency graph \code{G}, see Listings
\ref{lst:SIRrun} and \ref{lst:trSIR} in Appendix~\ref{sec:C-code-SIR}
for an example from the predefined \code{SIR} model in \pkg{SimInf}
and the use of the address of operator \code{'\&'} to obtain the
address of a function.  The \code{SimInf_model_run} function must
return the result from the call to the core solver with
\code{SimInf_run}.  The arguments to \code{SimInf_run} are the
arguments passed to the \code{SimInf_model_run} function plus the
vector of function pointers to the transition rate functions and the
function pointer to the post time step function. The add-on
\pkg{PredatorPrey} source package can now be built and installed with
the following commands.

\begin{Schunk}
\begin{Sinput}
R> pkg <- file.path(path, "PredatorPrey")
R> install.packages(pkg, repos = NULL, type = "source")
\end{Sinput}
\end{Schunk}

\begin{figure}
  \begin{center}
    \includegraphics{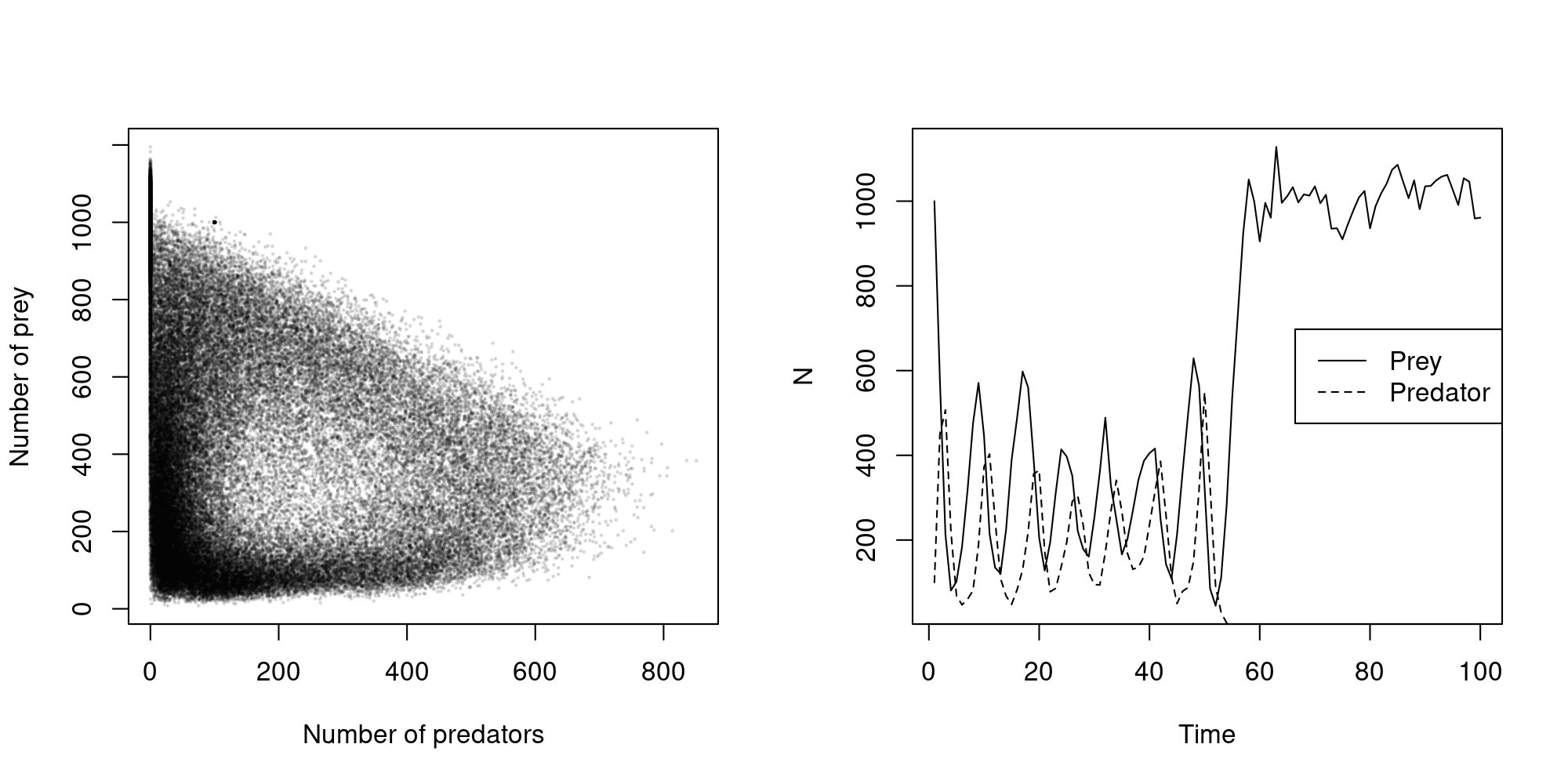}
  \end{center}
  \caption{Left: Phase plane trajectories from 1000 realizations of
    the Rosenzweig-MacArthur predator-prey model. Right: One
    realization of the Rosenzweig-MacArthur predator-prey model, where
    the predators go extinct and then the prey population fluctuates
    around a plateu of 1000 individuals.  \label{fig:predator-prey}}
\end{figure}

Here we let \code{repos = NULL} to install from local files and use
\code{type = "source"} to compile the files.  If the installation was
successful, the newly installed package \pkg{PredatorPrey} can be
loaded in \proglang{R} with the following command.

\begin{Schunk}
\begin{Sinput}
R> library("PredatorPrey")
\end{Sinput}
\end{Schunk}

Now create a model and run it to generate data.

\begin{Schunk}
\begin{Sinput}
R> model <- PredatorPrey(u0 = u0, tspan = 1:100, gdata = parameters)
R> set.seed(123)
R> result <- run(model, threads = 1)
\end{Sinput}
\end{Schunk}

Because a \code{PredatorPrey} object contains the \code{SimInf_model}
class, it can make use of all utility functions provided in the
\pkg{SimInf} package, for example, \code{show()}.

\begin{Schunk}
\begin{Sinput}
R> result
\end{Sinput}
\begin{Soutput}
Model: PredatorPrey
Number of nodes: 1000
Number of transitions: 5
Number of scheduled events: 0

Global data
-----------
 Parameter Value
 bR        2.0e+00
 bF        2.0e+00
 dR        1.0e+00
 K         1.0e+03
 alpha     7.0e-03
 w         3.5e-03
 dF        2.0e+00

Compartments
------------
   Min. 1st Qu. Median Mean 3rd Qu. Max.
 R    8     236    607  598     975 1195
 F    0       0     35  112     161  851
\end{Soutput}
\end{Schunk}

Or the \code{trajectory()} method, for example, to plot the phase
plane from 1000 realizations or to illustrate stochastic extinction of
the predators in the fourth node (Figure~\ref{fig:predator-prey}).

\begin{Schunk}
\begin{Sinput}
R> opar <- par(mfrow = c(1, 2))
R> plot(R ~ F, trajectory(model = result), cex = 0.3, pch = 20,
+    xlab = "Number of predators", ylab = "Number of prey",
+    col = rgb(0, 0, 0, alpha = 0.1))
R> plot(R ~ time, trajectory(model = result, node = 4), type = "l",
+    xlab = "Time", ylab = "N")
R> lines(F ~ time, trajectory(model = result, node = 4), type = "l", lty = 2)
R> legend("right", c("Prey", "Predator"), lty = 1:2)
R> par(opar)
\end{Sinput}
\end{Schunk}

This example illustrated how \pkg{SimInf} supports usage of the
numerical solvers from other \proglang{R} packages via the
\emph{LinkingTo} feature in \proglang{R}.


\section{Benchmark}
\label{sec:benchmark}

A comprehensive analysis of the performance of the numerical solver in
\pkg{SimInf} is presented by \citet{Bauer2016}.  Here, we provide a
small benchmark of the run-time of an SIR model using three
\proglang{R} packages on CRAN.  The measurements were obtained on a
ThinkPad T460p, Intel core i7-6700HQ quad-core at 2.6GHz , 8GB 2133MHz
RAM, running Fedora 27 and using \proglang{R} version 3.4.3.  Ten
replicates were performed and average run-time was estimated to
generate 1,000 realizations of an SIR model with parameters $\beta =
0.16$ and $\gamma = 0.077$ and initial conditions $S = 1000$, $I = 10$
and $R = 0$.  As shown in Table~\ref{table:benchmark}, the
implementation in \pkg{SimInf} appears to run over ten times faster
than \pkg{adaptivtau}.  This difference probably depends on
\pkg{adaptivetau} using a hybrid \proglang{R}/\proglang{C++}
implementation with \proglang{R} code for the transition rate
functions while \pkg{SimInf} uses \proglang{C} code.  To reduce
run-time further, \pkg{SimInf} has built-in support to perform
computations in parallel.  As expected, \pkg{GillespieSSA} has the
longest run-time since it has an implementation in pure \proglang{R}.

\begin{table*}[!ht]
  \small
  \center
  \begin{tabular}{l l c r}
    \toprule \proglang{R} package & Method & Threads & Time [ms]\\
    \midrule
    \pkg{SimInf} & Direct SSA & 4 &  57 \\
    \pkg{SimInf} & Direct SSA & 2 & 106 \\
    \pkg{SimInf} & Direct SSA & 1 & 198 \\
    \pkg{adaptivetau} & Tau-leaping & 1 & 2840 \\
    \pkg{adaptivetau} & Direct SSA & 1 & 9032 \\
    \pkg{GillespieSSA} & Tau-leaping & 1 & 30681 \\
    \pkg{GillespieSSA} & Direct SSA & 1 & 106091 \\
    \bottomrule
  \end{tabular}
  \caption{Comparison of the average run-time for generating 1000
    realizations of an SIR model.}
  \label{table:benchmark}
\end{table*}


\section{Conclusion}

In this paper we have introduced the \proglang{R} package \pkg{SimInf}
which supports data-driven simulations of disease transmission over
spatio-temporal networks. The package offers a very efficient and
highly flexible tool to incorporate real data in simulations at
realistic scales.

We hope that our package will facilitate incorporating available data,
for example, livestock data, in network epidemic models to better
understand disease transmission in a temporal network and improve
design of intervention strategies for endemic and emerging threats.
Future efforts will be concentrated on a software development driven
predominantly by actual use cases.


\section{Acknowledgments}

This work was financially supported by the Swedish Research Council
within the UPMARC Linnaeus centre of Excellence (P.~Bauer,
R.~Eriksson, S.~Engblom), the Swedish Research Council Formas
(S.~Engblom, S.~Widgren), the Swedish Board of Agriculture
(S.~Widgren), and by the Swedish strategic research program eSSENCE
(S.~Widgren).  The authors also thank two anonymous reviewers for
their helpful suggestions and comments, which greatly improved the
quality of both the software package and this manuscript.

\bibliography{SimInf}

\clearpage


\appendix

\section{Pseudo-code for the default simulation solver}
\label{sec:pseudo-code}

\begin{table*}[!ht]
  \small
  \begin{tabular}{r l}
    \toprule
    & \textbf{Algorithm} Pseudo-code for the default simulation solver
    using direct SSA \\
    \midrule

    \texttt{1:} & \textit{Run trajectory:} Dispatch to model specific
    \code{run} method. \\

    \texttt{2:} & \textit{\proglang{C} interface:} Initialize model
    transition rate functions and post time step function. \\

    \texttt{3:} & \textbf{for all} nodes $i=1$ \textbf{to} $\Nnodes$
    \textbf{do in parallel} \\

    \texttt{4:} & \quad Compute transition rates for all transitions
    $\omega_{i,j}$, $j=1, \ldots ,\Ntransitions$. \\

    \texttt{5:} & \textbf{end for} \\

    \texttt{6:} & \textbf{while} $t < T_{\text{End}}$ \textbf{do} \\

    \texttt{7:} & \quad \textbf{for all} nodes $i=1$ \textbf{to}
    $\Nnodes$ \textbf{do in parallel} \\

    \texttt{8:} & \quad \quad \textbf{loop} \\

    \texttt{9:} & \quad \quad \quad Compute sum of transition rates
    $\lambda_i = \sum_{j=1}^{\Ntransitions}\omega_{i,j}$ \\

    \texttt{10:} & \quad \quad \quad Sample time to next stochastic
    event $\tau_i = -\log(r_1)/ \lambda_i$ where $r_1$ \\ & \quad
    \quad \quad is a uniformly distributed random number in the range
    (0, 1) \\

    \texttt{11:} & \quad \quad \quad \textbf{if} $\tau_i + t_i >=
    T_{\text{Next day}}$ \textbf{then} \\

    \texttt{12:} & \quad \quad \quad \quad Move simulated time forward
    $t_i = T_{\text{Next day}}$ \\

    \texttt{13:} & \quad \quad \quad \quad go to \texttt{20} \\ 

    \texttt{14:} & \quad \quad \quad \textbf{end if} \\

    \texttt{15:} & \quad \quad \quad Move simulated time forward $t_i
    = t_i + \tau_i$ \\

    \texttt{16:} & \quad \quad \quad Determine which state transition
    happened; by inversion, \\ & \quad \quad \quad find $n$ such that
    $\sum_{j=1}^{n-1} \omega_{i,j} < \lambda r_2 \le \sum_{j=1}^n
    \omega_{i,j}$ where $r_2$ \\ & \quad \quad \quad is a uniformly
    distributed random number in the range (0, 1) \\

    \texttt{17:} & \quad \quad \quad Update the compartments \code{u[,
        i]} using the state-change vector \code{S[, n]} \\

    \texttt{18:} & \quad \quad \quad Use the dependency graph
    \code{G[, n]} to recalculate affected transition rates
    $\omega_{i,j}$ \\

    \texttt{19:} & \quad \quad \textbf{end loop} \\

    \texttt{20:} & \quad \quad Process $E_1$ events \\

    \texttt{21:} & \quad \textbf{end for} \\

    \texttt{22:} & \quad Process $E_2$ events \\

    \texttt{23:} & \quad \textbf{for all} nodes $i=1$ \textbf{to}
    $\Nnodes$ \textbf{do in parallel} \\

    \texttt{24:} & \quad \quad Call post time step function and update
    the continuous state variable \code{v[ ,i]}. \\

    \texttt{25:} & \quad \textbf{end for} \\

    \texttt{26:} & \quad $T_{\text{Next day}} = T_{\text{Next day}} +
    1$ \\

    \texttt{27:} & \textbf{end while} \\

    \bottomrule
  \end{tabular}
\end{table*}

\clearpage

\section{Illustration of scheduled events}
\label{sec:illustration-scheduled-events}

\begin{figure}[!h]
  \begin{center}
    \includegraphics[width=0.67\linewidth]{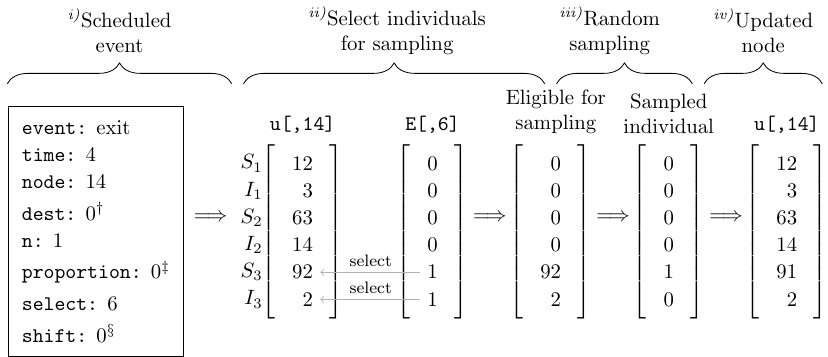}
  \end{center}
  \caption{Illustration of a scheduled \textit{exit} event in the
    \code{SISe3\_sp} model at time = 4.  The removal of one individual
    in the third age category $\{S_3, I_3\}$ from node 14.
    Interpreting the figure from left to right: \textit{i)} A single
    row of the event data operating on node 14.  \textit{ii)}
    \code{u[, 14]} is the current state of node 14; \code{E[, 6]} is
    the 6$^{th}$ column in the select matrix that determines which
    compartments (age categories) that are eligible for sampling.
    \textit{iii)} The operation of randomly sampling one individual (n
    = 1) to move from the compartments selected in step \textit{ii}.
    \textit{iv)} The resultant state of node 14 after subtracting the
    sampled individual in step \textit{iii} from node 14.
    $^\dag$\code{dest} and $^\S$\code{shift} are not used in a
    scheduled \textit{exit} event.  $^\ddag$\code{proportion} is not
    used when $n > 0$.  \label{fig:exit}}
\end{figure}

\begin{figure}[!h]
  \begin{center}
    \includegraphics[width=0.6\linewidth]{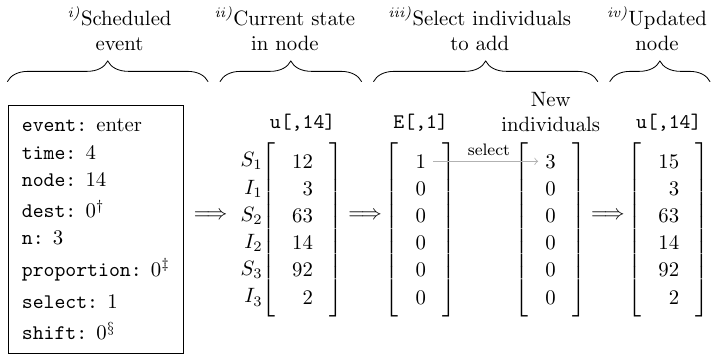}
  \end{center}
  \caption{Illustration of a scheduled \textit{enter} event in the
    \code{SISe3\_sp} model at time = 4.  Add three susceptible
    individuals to the first age category $\{S_1\}$ in node 14.
    Interpreting the figure from left to right: \textit{i)} A single
    row of the event data operating on node 14.  \textit{ii)}
    \code{u[, 14]} is the current state of node 14.  \textit{iii)}
    \code{E[, 1]} is the first column in the select matrix that
    determines which compartments (age categories) the new individuals
    are added.  \textit{iv)} The resultant state of node 14 after
    adding the individuals in step \textit{iii}.  $^\dag$\code{dest},
    $^\ddag$\code{proportion} and $^\S$\code{shift} are not used in a
    scheduled \textit{enter} event.  \label{fig:enter}}

  \vspace*{\floatsep}

  \begin{center}
    \includegraphics[width=0.66\linewidth]{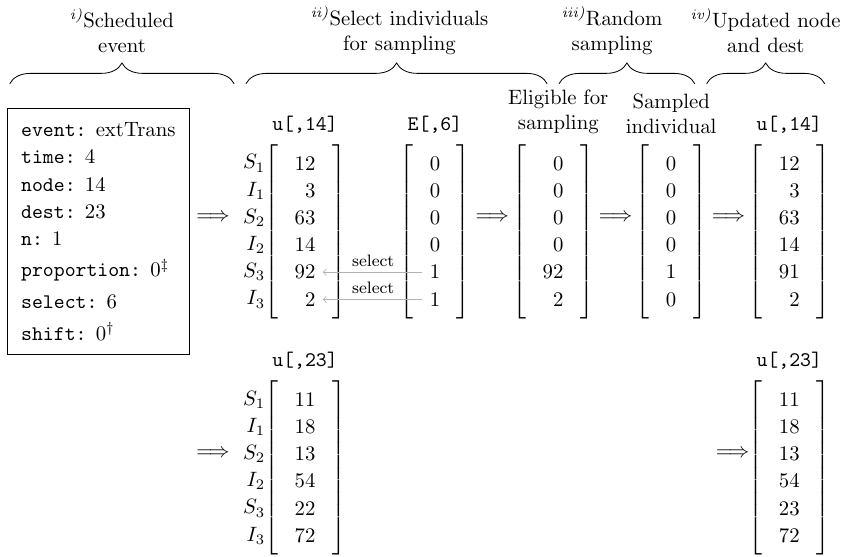}
  \end{center}
  \caption{Illustration of a scheduled \textit{external transfer}
    event in the \code{SISe3\_sp} model at time = 4.  The movement of
    one individual in the third age category $\{S_3, I_3\}$ from node
    14 to destination node 23.  Interpreting the figure from left to
    right: \textit{i)} A single row of the event data operating on
    node 14 and destination node 23.  \textit{ii)} \code{u[, 14]} is
    the current state of node 14; \code{u[, 23]} is the current state
    of the destination node 23; \code{E[, 6]} is the 6$^{th}$ column
    in the select matrix that determines which compartments (age
    categories) that are eligible for sampling.  \textit{iii)} The
    operation of randomly sampling one individual (n = 1) to move from
    the compartments selected in step \textit{ii}.  \textit{iv)} The
    resultant state of node 14 and destination node 23 after
    subtracting the sampled individuals in step \textit{iii} from node
    14 and adding them to destination node 23.  $^\dag$\code{shift}
    can be used in a scheduled \textit{external transfer} event, see
    Figure~\ref{fig:external:shift}. $^\ddag$\code{proportion} is not
    used when $n > 0$.  \label{fig:external}}
\end{figure}

\begin{figure}[!h]
  \begin{center}
    \includegraphics[width=0.8\linewidth]{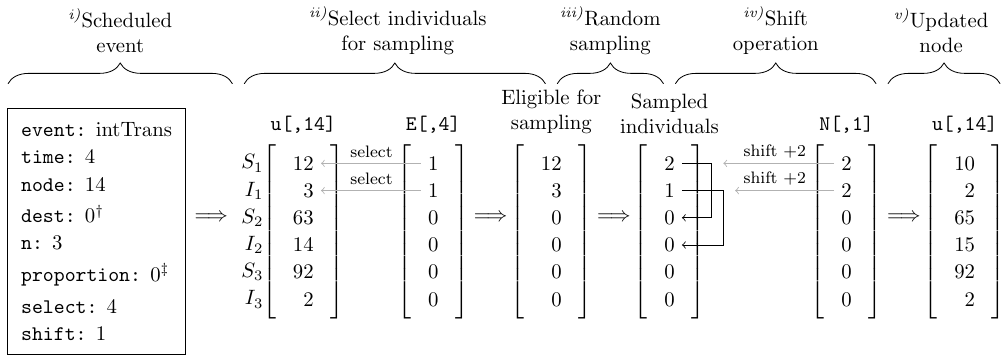}
  \end{center}
  \caption{Illustration of a scheduled \textit{internal transfer}
    event in the \code{SISe3\_sp} model at time = 4.  The ageing of
    three individuals in the first age category $\{S_1, I_1\}$.
    Interpreting the figure from left to right: \textit{i)} A single
    row of the event data operating on node 14.  \textit{ii)}
    \code{u[, 14]} is the current state of node 14; \code{E[, 4]} is
    the 4$^{th}$ column in the select matrix that determines which
    compartments (age categories) that are eligible for sampling.
    \textit{iii)} The operation of randomly sampling three individuals
    (n = 3) to age from the compartments selected in step \textit{ii}.
    \textit{iv)} The shift operation applies the shift specified in
    column 1 of the shift matrix (\code{N}) to the individuals sampled
    in step \textit{iii}.  \textit{v)} The resultant state of node 14
    after subtracting the sampled individuals in step \textit{iii} and
    adding the individuals after the shift operation in step
    \textit{iv}.  $^\dag$\code{dest} is not used in \textit{internal
      transfers}.  $^\ddag$\code{proportion} is not used when $n >
    0$.  \label{fig:internal}}

  \vspace*{\floatsep}

  \begin{center}
    \includegraphics[width=0.66\linewidth]{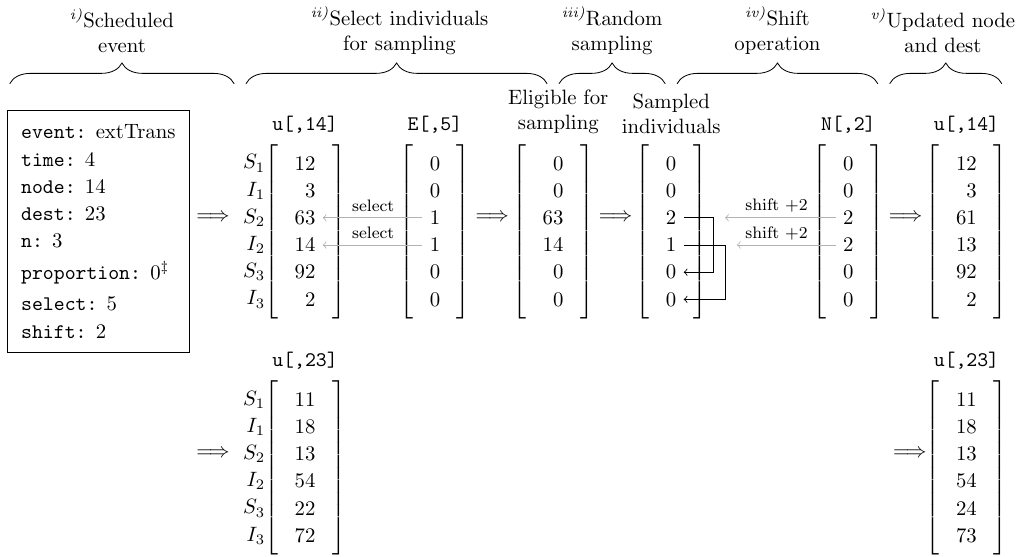}
  \end{center}
  \caption{Illustration of a scheduled \textit{external transfer}
    event in the \code{SISe3} model at time = 4.  The ageing of three
    individuals in the second age category $\{S_2, I_2\}$ that are
    subsequently moved.  Interpreting the figure from left to right:
    \textit{i)} A single row of the event data operating on node 14
    and destination node 23.  \textit{ii)} \code{u[, 14]} is the
    current state of node 14; \code{u[, 23]} is the current state of
    the destination node 23; \code{E[, 5]} is the 5$^{th}$ column in
    the select matrix that determines which compartments (age
    categories) that are eligible for sampling.  \textit{iii)} The
    operation of randomly sampling three individuals (n = 3) to move
    from the compartments selected in step \textit{ii}.  \textit{iv)}
    The shift operation applies the shift specified in column 2 of the
    shift matrix (\code{N}) to the individuals sampled in step
    \textit{iii}.  \textit{v)} The resultant state of node 14 and
    destination node 23 after subtracting the sampled individuals in
    step \textit{iii} from node 14 and adding them to the destination
    node 23 after the shift operation in step \textit{iv}.
    $^\ddag$\code{proportion} is not used when $n >
    0$.  \label{fig:external:shift}}
\end{figure}

This section illustrates how the scheduled events for the
\code{SISe3\_sp} model are specified
(Table~\ref{table:SISe3_sp:events}) and how each event type is
executed
Figures~\ref{fig:exit},\ref{fig:enter},\ref{fig:external},\ref{fig:internal},\ref{fig:external:shift}

\begin{table}[!ht]
  \small
  \begin{tabular}{l c c c c c c c c}
    \toprule
    Action & \code{event} & \code{time} & \code{node} & \code{dest} &
    \code{n} & \code{proportion} & \code{select} & \code{shift} \\
    \midrule
    Exit individuals in $S_1$ and $I_1$  & exit     & t & i & 0 & n & 0 & 4 & 0\\
    Exit individuals in $S_2$ and $I_2$  & exit     & t & i & 0 & n & 0 & 5 & 0\\
    Exit individuals in $S_3$ and $I_3$  & exit     & t & i & 0 & n & 0 & 6 & 0\\
    Enter individuals in $S_1$ and $I_1$ & enter    & t & i & 0 & n & 0 & 1 & 0\\
    Enter individuals in $S_2$ and $I_2$ & enter    & t & i & 0 & n & 0 & 2 & 0\\
    Enter individuals in $S_3$ and $I_3$ & enter    & t & i & 0 & n & 0 & 3 & 0\\
    Age individuals in $S_1$ and $I_1$   & intTrans & t & i & 0 & n & 0 & 4 & 1\\
    Age individuals in $S_2$ and $I_2$   & intTrans & t & i & 0 & n & 0 & 5 & 2\\
    Move individuals in $S_1$ and $I_1$  & extTrans & t & i & j & n & 0 & 4 & 0\\
    Move individuals in $S_2$ and $I_2$  & extTrans & t & i & j & n & 0 & 5 & 0\\
    Move individuals in $S_3$ and $I_3$  & extTrans & t & i & j & n & 0 & 6 & 0\\
    \bottomrule
  \end{tabular}
  \caption{Examples of the specification of a single row of scheduled
    event data in the \code{SISe3\_sp} model to add, move or remove
    individuals during the simulation, where $t$ is the time-point for
    the event, $i$ is the node to operate on, $j$ is the destination
    node for a movement.}
  \label{table:SISe3_sp:events}
\end{table}

\clearpage

\section[C code for the SIR model]{\proglang{C} code for the \code{SIR} model}
\label{sec:C-code-SIR}

\begin{minipage}{\linewidth}
\begin{scriptsize}
  \begin{lstlisting}[
      language=C, caption={Implementation of the function to init and
        run a simulation with the \code{SIR} model},
      label={lst:SIRrun}, frame=single]
  SEXP SIR_run(SEXP model, SEXP threads, SEXP solver)
  {
      TRFun tr_fun[] = {&SIR_S_to_I, &SIR_I_to_S};

      return SimInf_run(model, threads, solver, tr_fun, &SIR_post_time_step);
  }
\end{lstlisting}
\end{scriptsize}
\end{minipage}

\begin{minipage}{\linewidth}
\begin{scriptsize}
\begin{lstlisting}[
    language=C, caption={Implementation of the transition rate
      functions in the \code{SIR} model for the transitions in
      Equation~\ref{eq:SIR} between the susceptible and infected
      compartments.  The enumeration declarations are used to name the
      variable offsets and facilitate extraction of the values from
      the various data vectors.}, label={lst:trSIR} , frame=single]

  /* Offset in integer compartment state vector */
  enum {S, I, R};

  /* Offsets in global data (gdata) to parameters in the model */
  enum {BETA, GAMMA};

  /* susceptible to infected: S -> I */
  double SIR_S_to_I(const int *u, const double *v, const double *ldata,
                    const double *gdata, double t)
  {
      const double S_n = u[S];
      const double I_n = u[I];
      const double n = S_n + I_n + u[R];

      if (n > 0.0)
          return (gdata[BETA] * S_n * I_n) / n;
      return 0.0;
  }

  /* infected to susceptible: I -> S */
  double SIR_I_to_S(const int *u, const double *v, const double *ldata,
                    const double *gdata, double t)
  {
      return gdata[GAMMA] * u[I];
  }
\end{lstlisting}

\begin{lstlisting}[
      language=C, caption={Implementation of the post time step
        function in the \code{SIR} model.  The post time step function
        should return a value $>0$ if the node needs to recalculate
        the transition rates for the node, a value (error code) $<0$
        if an error is detected, or otherwise $0$.  Since the post
        time step function for the \code{SIR} model does not make any
        changes to a node, it always return $0$.}, label={lst:ptsSIR}
      ,frame=single]

  int SIR_post_time_step(double *v_new, const int *u, const double *v,
                         const double *ldata, const double *gdata,
                         int node, double t)
  {
      return 0;
  }
\end{lstlisting}
\end{scriptsize}
\end{minipage}

\end{document}